\documentclass[aps,pra,amsmath,amssymb,twocolumn,superscriptaddress,longbibliography]{revtex4-1}
\usepackage{ulem}
\usepackage{amsmath,amssymb}
\usepackage{graphicx}	
\def\vecb{\boldsymbol}

\begin{document}

\author{Shunsuke~A.~Sato}
\email{shunsuke.sato@tohoku.ac.jp}
\affiliation 
{Department of Physics, Tohoku University, Sendai 980-8578, Japan}
\affiliation{Max Planck Institute for the Structure and Dynamics of Matter and Center for Free Electron Laser Science, 22761 Hamburg, Germany}

\author{Hannes~H\"ubener}
\affiliation{Max Planck Institute for the Structure and Dynamics of Matter and Center for Free Electron Laser Science, 22761 Hamburg, Germany}

\author{Umberto~De~Giovannini}
\affiliation{Universit\`a degli Studi di Palermo, Dipartimento di Fisica e Chimica—Emilio Segr\`e, via Archirafi 36, I-90123 Palermo, Italy}
\affiliation{Max Planck Institute for the Structure and Dynamics of Matter and Center for Free Electron Laser Science, 22761 Hamburg, Germany}

\author{Angel~Rubio}
\affiliation{Max Planck Institute for the Structure and Dynamics of Matter and Center for Free Electron Laser Science, 22761 Hamburg, Germany}
\affiliation 
{Initiative for Computational Catalysis, Flatiron Institute, 162 Fifth Avenue, New York, NY 10010, USA}

\title{Polarized Houston State Framework for Nonequilibrium Driven Open Quantum Systems}

\begin{abstract}
We introduce a new theoretical framework --the polarized Houston basis-- to model nonequilibrium dynamics in driven open quantum systems, formulated for use within the quantum master equation. This basis extends conventional Houston states by incorporating field-induced polarization effects, enabling a more accurate description of excitation dynamics under external driving. Using a one-dimensional dimer-chain model, we examine band population dynamics through projections onto polarized Houston states, original Houston states, and naive Bloch states. We find that the polarized Houston basis significantly suppresses spurious Bloch-state excitations and virtual transitions present in standard Houston approaches, allowing for a cleaner extraction of real excitations. When implemented in the relaxation time approximation of the quantum master equation, this formalism also yields a substantial reduction of unphysical DC currents in insulating systems. Our results highlight the polarized Houston basis as a powerful tool for simulating nonequilibrium phenomena in light-driven open quantum materials.
\end{abstract}

\maketitle

\section{Introduction \label{sec:intro}}

The band population of nonequilibrium systems is an important quantity for understanding dynamical processes in condensed matter. For example, the band population dynamics after optical pumping has been investigated using time-resolved angle-resolved photoemission spectroscopy (Tr-ARPES). The population dynamics in momentum space provide information about the relaxation paths and timescales of photo-excited carriers~\cite{PhysRevLett.115.086803,PhysRevLett.116.076801,PhysRevLett.117.277201,10.1063/1.4965839,Roth_2019}. Furthermore, field-induced photoexcitation processes have been revealed on the attosecond timescale using attosecond transient absorption spectroscopy~\cite{Schultze2013,doi:10.1126/science.1260311,Zurch2017,Schlaepfer2018,Inzani2023}.

Theoretically, in static or equilibrium systems, band populations can be evaluated by projecting wavefunctions onto static Bloch states, which are the eigenstates of the unperturbed Hamiltonian. However, in the presence of external driving fields, the system enters a dynamically evolving regime, and the choice of projection basis becomes nontrivial. A naive projection onto static Bloch states can introduce spurious population dynamics, as it fails to account for field-induced intraband motion. To address this, projections onto Houston states, which incorporate intraband acceleration due to the field, have been widely adopted as a more physically appropriate alternative. These states offer a cleaner representation of population dynamics under external fields by eliminating artificial excitations associated with the static Bloch picture~\cite{PhysRevB.77.165104}.

By its nature, projection onto Houston states may result in dynamical populations under the fields that include virtual excitations, which occur only during field irradiation and disappear once the fields are removed. While virtual excitation itself may provide insightful information for investigating light-induced phenomena~\cite{McDonald_2017,PhysRevB.102.041125,Boolakee2022,PhysRevB.107.075135}, it is often inconvenient as it obscures the real excitation dynamics~\cite{doi:10.1126/science.1260311}. Therefore, for studies of photocarrier injection processes, it is desirable to develop methods for analyzing population dynamics without the contribution of virtual excitation. Moreover, in the context of open quantum systems, it has been reported that the relaxation time approximation with Houston states can cause spurious DC responses in insulators under static fields, due to the lack of field-induced modifications in electronic states~\cite{PhysRevB.109.L180302}. This highlights the need for a set of reference states that incorporate field-induced effects to naturally describe the electronic structure under external fields.

In this work, we introduce the polarized Houston basis as a generalization of conventional Houston states that accounts for field-induced polarization effects, providing a more physically consistent framework for modeling nonequilibrium dynamics in driven open quantum systems. Building on this foundation, we develop a projection method for evaluating band population dynamics under external gauge fields. This approach yields population distributions that naturally reflect real carrier injection processes, free from the artifacts seen in naive Bloch or standard Houston representations. We further implement the polarized Houston basis within the relaxation time approximation of the quantum master equation, and demonstrate its effectiveness by evaluating the DC current response of a model insulator. Our results show that the unphysical DC currents often introduced by conventional Houston-state treatments are completely eliminated, underscoring the utility of polarized Houston states for open-system quantum transport modeling.

The paper is organized as follows: In Sec.~II, we introduce polarized Houston states for describing reference states in the presence of time-dependent gauge fields. We also discuss static Bloch states and Houston states from the perspective of reference states in dynamical systems. In Sec.~III, we evaluate the dynamical population using projections onto static Bloch, Houston, and polarized Houston states under various conditions, employing the time-dependent Schr\"odinger equation. In Sec.~IV, we investigate polarized Houston states as reference states within the relaxation time approximation in the framework of the quantum master equation. Finally, our findings are summarized in Sec.~V.

\section{Transient states of time-dependent driven quantum systems \label{sec:projection}}

In this section, we explore several candidate basis sets that can serve as projectors for analyzing dynamical band populations in driven solid-state systems. For clarity and simplicity, we restrict our discussion to electronic dynamics governed by the following one-body time-dependent Schr\"odinger equation:
\begin{align}
i\hbar \frac{\partial}{\partial t}u_{b\vecb k}(\vecb r, t) &= \left [\frac{\left (\vecb p + \hbar \vecb k + e\vecb A(t) \right )^2}{2m} + v(\vecb r) \right ]u_{b\vecb k}(\vecb r,t) \nonumber \\
&= \hat h_{\vecb k + e\vecb A(t)/\hbar}u_{b\vecb k}(\vecb r,t),
\label{eq:tdse}
\end{align}  
where $ u_{b\vecb k}(\vecb r, t) $ is the periodic part of the time-dependent Bloch state, $b$ is the band index, and $\vecb k$ is the Bloch wavevector. The Bloch wavefunction satisfies $u_{b\vecb k}(\vecb r + \vecb a, t) = u_{b\vecb k}(\vecb r, t)$, where $\vecb a$ represents the lattice vectors. The external driving field is described by the homogeneous vector potential $\vecb A(t)$ under the dipole approximation, and the one-body potential $v(\vecb r)$ has the same periodicity of the lattice; $v(\vecb r + \vecb a) = v(\vecb r)$. The one-body Hamiltonian can be expressed as $\hat h_{\vecb k + e\vecb A(t)/\hbar}$, as the time dependence manifests only through the wavevector shift $\vecb k \to \vecb k + e\vecb A(t)/\hbar$.

By analyzing the time-dependent Bloch states $u_{b\vecb k}(\vecb r, t)$, the dynamical population of each band may be evaluated using projections onto a chosen basis set. One of the simplest choices is the set of eigenstates of the field-free Hamiltonian, defined by  
\begin{align}
\hat h_{\vecb k}u^{\mathrm{B}}_{b\vecb k}(\vecb r) = \epsilon_{b\vecb k}u^{\mathrm{B}}_{b\vecb k}(\vecb r).
\label{eq:bloch-basis}
\end{align}  
We refer to this as the \textit{Bloch basis}. The dynamical population of a band can then be determined based on its projection onto the Bloch basis:  
\begin{align}
n^{\mathrm{B}}_{b\vecb k}(t) &= \sum_{b'} \left | \int_{\Omega} d\vecb r \left ( u^{\mathrm{B}}_{b\vecb k}(\vecb r) \right )^* u_{b'\vecb k}(\vecb r, t) \right |^2 \nonumber \\
&= \sum_{b'} \left | \langle u^{\mathrm{B}}_{b\vecb k} | u_{b'\vecb k}(t) \rangle \right |^2,
\end{align}  
where $\Omega$ denotes the volume of the unit cell.

Although the Bloch states, $|u^{\mathrm{B}}_{b\vecb{k}}\rangle$, and the resulting population, $n^{\mathrm{B}}_{b\vecb{k}}(t)$, can be easily evaluated numerically, it becomes challenging to evaluate the population in the presence of a finite vector potential, $\vecb{A}(t)$~\cite{PhysRevB.77.165104}. This difficulty arises because the Bloch states do not account for the intraband motion caused by the wavevector shift, $\vecb{k} \rightarrow \vecb{k} + e\vecb{A}(t)/\hbar$.

An alternative choice for the projection basis is the \textit{Houston basis}~\cite{PhysRev.57.184,PhysRevB.33.5494}. The Houston states are the instantaneous eigenstates of the Hamiltonian, $h_{\vecb{k} + e\vecb{A}(t)/\hbar}$. To introduce the Houston states, we first consider the adiabatic solution of the time-dependent Schr\"odinger equation, Eq.~(\ref{eq:tdse}), as follows:
\begin{align}
u^{\mathrm{H}}_{b\vecb{k}}(\vecb{r}, t) &= \exp \left[\frac{1}{i\hbar} \int_{-\infty}^t dt' \, \epsilon_{b, \vecb{k} + e\vecb{A}(t')/\hbar} \right] \nonumber \\
&\quad \times 
\exp \left[i \int^{t}_{-\infty}dt' \frac{e}{\hbar} \dot{\vecb A}(t') \cdot  \vecb{A}^{\mathrm{BC}}_b\left(\vecb{k} + \frac{e}{\hbar} \vecb{A}(t') \right) \right] \nonumber \\
&\quad \times u^{\mathrm{B}}_{b, \vecb{k} + e\vecb{A}(t)/\hbar}(\vecb{r}),
\label{eq:houston-states}
\end{align}
where $\vecb{A}^{\mathrm{BC}}_b\left(\vecb{k} \right)$ is the Berry connection, defined as
\begin{align}
\vecb{A}^{\mathrm{BC}}_{b}\left(\vecb{k} \right) = i \langle u^{\mathrm{B}}_{b\vecb{k}}| \frac{\partial}{\partial \vecb k} |u^{\mathrm{B}}_{b\vecb{k}}\rangle.
\end{align}
Here, we note that the following geometric phase depends only on the path of a parameter $\vecb K$ as
\begin{align}
&\exp \left[i \int^{t}_{-\infty}dt' \frac{e}{\hbar} \dot{\vecb A}(t') \cdot  \vecb{A}^{\mathrm{BC}}_b\left(\vecb{k} + \frac{e}{\hbar} \vecb{A}(t') \right) \right] \nonumber \\
&=\exp \left[i \int^{\vecb K(t)}_{\vecb K(-\infty)}d \vecb K \cdot  \vecb{A}^{\mathrm{BC}}_b\left(\vecb{K} \right) \right],
\end{align}
where $\vecb K(t)$ is defined as
\begin{align}
\vecb K(t) = \vecb{k} + \frac{e}{\hbar} \vecb{A}(t).
\end{align}

In this work, we refer to the adiabatic solution $u^{\mathrm{H}}_{b\vecb{k}}(\vecb{r}, t)$ as the \textit{Houston states}, and the set of these states as the \textit{Houston basis}. Employing the Houston basis for band population analysis eliminates artificial population dynamics induced by the gauge field~\cite{PhysRevB.77.165104}. Thus, the Houston basis is a natural choice for projecting and analyzing the instantaneous band population in the presence of external fields.

It is worth noting that the Houston states are instantaneous eigenstates of the time-dependent Hamiltonian, $\hat{h}_{\vecb{k} + e\vecb{A}(t)/\hbar}$, in the velocity gauge expression. They correspond to the static eigenstates of the field-free Hamiltonian in the length gauge expression (see Appendix~\ref{appendix:sec:ex-hydrogen} and \ref{appendix:sec:ex-solids}). As we will demonstrate later, projections onto the Houston basis inherently include virtual excitation components, which arise from its treatment of intraband motion without accounting for field-induced polarization effects. In the analysis of driven systems, these virtual excitations can complicate the interpretation of real carrier injection, especially when the primary interest is in actual excitation dynamics. To mitigate this issue, we introduce the polarized Houston basis, derived from the polarized Houston states, which incorporates field-induced polarization and effectively suppresses the contribution of virtual excitations in population analysis.

To introduce the polarized Houston states, we first introduce the instantaneous eigenstates of the Hamiltonian, $\hat{h}_{\vecb{k} + e\vecb{A}(t)/\hbar}$, with the geometric phase, as follows:
\begin{align}
u^{\mathrm{A}}_{b\vecb{k}}(\vecb{r},t) &= 
\exp \left[i \int^{t}_{-\infty}dt' \frac{e}{\hbar} \dot{\vecb A}(t') \cdot  \vecb{A}^{\mathrm{BC}}_b\left(\vecb{k} + \frac{e}{\hbar} \vecb{A}(t') \right) \right] \nonumber \\
&\quad \times u^{\mathrm{B}}_{b, \vecb{k} + e\vecb{A}(t)/\hbar}(\vecb{r}).
\label{eq:adiabatic-basis}
\end{align}
Here, $u^{\mathrm{A}}_{b\vecb{k}}(\vecb{r},t)$ are identical to the Houston states, $u^{\mathrm{H}}_{b\vecb{k}}(\vecb{r},t)$, in Eq.~(\ref{eq:houston-states}), up to the dynamical phase factor. Thus, the states $u^{\mathrm{A}}_{b\vecb{k}}(\vecb{r},t)$ are uniquely defined geometrically in the Brillouin zone, or $\vecb{k}$-space, with respect to the shifted $k$-vector, $\vecb{k} + e\vecb{A}(t)/\hbar$.

We then expand the time-dependent Bloch states, $u_{b \vecb{k}}(\vecb{r},t)$, using the instantaneous eigenbasis:
\begin{align}
u_{b \vecb{k}}(\vecb{r},t) = \sum_{b'} c_{bb', \vecb{k}}(t) u^{\mathrm{A}}_{b'\vecb{k}}(\vecb{r},t),
\label{eq:houston-expansion}
\end{align}
where $c_{bb', \vecb{k}}(t)$ are the expansion coefficients. For convenience, we define a coefficient vector as $\vecb{c}_{b\vecb{k}}(t) = \left (c_{b, 1, \vecb{k}}(t), c_{b, 2, \vecb{k}}(t), \cdots \right )^\mathrm{T}$. Substituting Eq.~(\ref{eq:houston-expansion}) into Eq.~(\ref{eq:tdse}), we obtain the following equation of motion for the coefficient vector:
\begin{align}
i\hbar \frac{d}{dt}\vecb{c}_{b\vecb{k}}(t) = \mathcal{H}_{\mathrm{eff},\vecb{k}}(t)\vecb{c}_{b\vecb{k}}(t),
\label{eq:effective-tdse-adiabatic-basis}
\end{align}
where the effective Hamiltonian matrix is defined elementwise as:
\begin{align}
\left ( \mathcal{H}_{\mathrm{eff},\vecb{k}}(t) \right )_{bb'} &= \epsilon_{b, \vecb{k} + e\vecb{A}(t)/\hbar} \delta_{bb'} \nonumber \\
& + i \left (1 - \delta_{bb'} \right ) \frac{e\vecb{E}(t)}{\hbar} \cdot \langle u^{\mathrm{A}}_{b\vecb{k}}(\vecb{r},t) | \frac{\partial}{\partial \vecb{k}} | u^{\mathrm{A}}_{b'\vecb{k}}(\vecb{r},t) \rangle.
\label{eq:effective-ham-matrix}
\end{align}
By interpreting the following quantities as dipole matrix elements,
\begin{align}
\vecb d_{ab,\vecb k}(t) = \frac{i}{\hbar}
\langle u^{\mathrm{A}}_{b\vecb{k}}(\vecb{r},t) | \frac{\partial}{\partial \vecb{k}} | u^{\mathrm{A}}_{b'\vecb{k}}(\vecb{r},t) \rangle,
\end{align}
the effective Hamiltonian of Eq.~(\ref{eq:effective-ham-matrix}) can be understood as a combination of the diagonal elements, which describe the single-particle energies with the intraband shift, and the off-diagonal elements, which describe the interband transition via the dipole transition.

We then introduce the eigenvectors $\vecb{c}^P_{b\vecb k}(t)$ and eigenvalues $\epsilon^P_{b\vecb k}(t)$ of the Hamiltonian matrix as
\begin{align}
\mathcal{H}_{\mathrm{eff},\vecb k}(t) \vecb{c}^P_{b\vecb k}(t) = \epsilon^P_{b\vecb k}(t) \vecb{c}^P_{b\vecb k}(t).
\label{eq:effective-ham-in-length-gauge}
\end{align}

Using the elements of the eigenvectors, $\vecb{c}^P_{b\vecb k}(t) = \left(c^P_{b,1,\vecb k}(t), c^P_{b,2,\vecb k}(t), \cdots \right)^T$, we define the \textit{polarized basis states} as
\begin{align}
u^{\mathrm{P}}_{b\vecb k}(\vecb r,t) = \sum_{b'} c^P_{bb',\vecb k}(t) u^{\mathrm{A}}_{b'\vecb k}(\vecb r,t).
\end{align}
We further assign $\epsilon^P_{b\vecb k}(t)$ to the corresponding single-particle energy of the polarized basis states. Finally, we define the \textit{polarized Houston states} using the polarized basis states as
\begin{align}
u^{\mathrm{PH}}_{b\vecb k}(\vecb r,t) &=e^{i\gamma^P_{b\vecb k}(t)} \exp \left[\frac{1}{i\hbar} \int^t_{-\infty} dt' \, \epsilon^P_{b, \vecb k}(t') \right]  u^{\mathrm{P}}_{b, \vecb k}(\vecb r,t),
\label{eq:polarized-houston-states}
\end{align}
where $\gamma^P_{b\vecb k}(t)$ is the corresponding geometric phase, defined by
\begin{align}
\gamma^P_{b\vecb k}(t)= i\int^{t}_{-\infty}dt' \, \vecb{c}^{P,\dagger}_{b\vecb k}(t')
\dot{\vecb{c}}^P_{b\vecb k}(t').
\end{align}

We note that the Hamiltonian, Eq.~(\ref{eq:effective-ham-in-length-gauge}), is equivalent to the Hamiltonian in the length gauge with interband transitions~\cite{PhysRevB.52.14636,PhysRevB.96.195413}. Hence, the polarized basis states, \( u^{\mathrm{P}}_{b, \vecb{k}}(\vecb{r},t) \), and the polarized Houston states, \( u^{\mathrm{PH}}_{b\vecb{k}}(\vecb{r},t) \), are expected to capture aspects of field-induced polarization effects, such as the Stark effect.

In this section, we have introduced three kinds of quantum states to describe the dynamics of driven quantum systems in general. The first is the Bloch states, $u^{\mathrm{B}}_{b\vecb k}(\vecb r)$, as defined in Eq.~(\ref{eq:bloch-basis}). The second is the Houston states, $u^{\mathrm{H}}_{b\vecb k}(\vecb r,t)$, as defined in Eq.~(\ref{eq:houston-states}). The third is the polarized Houston states, $u^{\mathrm{PH}}_{b\vecb k}(\vecb r,t)$, as defined in Eq.~(\ref{eq:polarized-houston-states}). In the following sections, we will examine how these states capture the dynamical nature of the driven quantum system by evaluating the dynamical population using the time-dependent Schr\"odinger equation in Sec.~\ref{sec:pop-analysis-tdse} and the quantum master equation in Sec.~\ref{sec:ref-state-q-master}.

\section{Dynamical Band Population Analysis with the Exact Time-Dependent Schr\"odinger Equation \label{sec:pop-analysis-tdse}}

In this section, we examine three types of states, $u^{\mathrm{B}}_{b\vecb k}(\vecb r)$, $u^{\mathrm{H}}_{b\vecb k}(\vecb r, t)$, and $u^{\mathrm{PH}}_{b\vecb k}(\vecb r, t)$, as projectors for evaluating the dynamical band population. To proceed with the practical analysis, we consider a one-dimensional dimer-chain model described by the following Hamiltonian:  
\begin{align}  
H_{k} = \begin{pmatrix}  
-\frac{\Delta}{2} & -2t_H \cos \left [ \frac{a_L}{2} k \right ] \\  
-2 t_H \cos \left [ \frac{a_L}{2} k \right ] & \frac{\Delta}{2}  
\end{pmatrix},  
\label{eq:ham-dimer-chain}  
\end{align}  
where $k$ is the Bloch wavenumber, $\Delta$ is the bandgap, $t_H$ is the hopping energy, and $a_L$ is the lattice constant. In this work, we choose the parameters to reproduce several electronic properties of GaAs. We set $a_L$ to $5.65$~\AA and $\Delta$ to $1.52$~eV~\cite{10.1063/1.1368156}. Furthermore, we set $t_H$ to 1.58~eV so as to reproduce the electron-hole reduced mass, $1/m^{*}=1/m_{\mathrm{CB}}+1/m_{m_{lh,\mathrm{VB}}}$ with the conduction electron mass ($m_{\mathrm{CB}}=0.067m_e$~\cite{10.1063/1.1368156}) and the valence light-hole mass ($m_{lh, \mathrm{VB}}=0.08m_e$~\cite{peter2010fundamentals}).

We introduce the eigenstates $\vecb{u}^{\mathrm{B}}_{bk}$ and the eigenvalues $\epsilon_{bk}$ of the Hamiltonian in Eq.~(\ref{eq:ham-dimer-chain}) as  
\begin{align}  
H_{k} \vecb{u}^{\mathrm{B}}_{bk} = \epsilon_{bk} \vecb{u}^{\mathrm{B}}_{bk},  
\end{align}  
where $b$ denotes the band index, specifying the valence band ($b = v$) or the conduction band ($b = c$). The eigenstates, $\vecb{u}^{\mathrm{B}}_{bk}$, correspond to the Bloch states $u^{\mathrm{B}}_{b\vecb k}(\vecb r)$ defined in Eq.~(\ref{eq:bloch-basis}).  

To describe field-induced electron dynamics, we set the valence Bloch state as the initial condition for the time-dependent problem and solve the following time-dependent Schr\"odinger equation:  
\begin{align}  
i\hbar \frac{d}{dt} \vecb{u}_{k}(t) = H_{k+eA(t)/\hbar} \vecb{u}_{k}(t),  
\label{eq:tdse-dimer-chain}  
\end{align}  
where $A(t)$ is the time-dependent vector potential.  

Based on Eq.~(\ref{eq:tdse-dimer-chain}), we introduce the Houston states, $\vecb{u}^{\mathrm{H}}_{bk}(t)$, and the polarized Houston states, $\vecb{u}^{\mathrm{PH}}_{bk}(t)$, as defined by Eq.~(\ref{eq:houston-states}) and Eq.~(\ref{eq:polarized-houston-states}), respectively. We then define the dynamical conduction population at each time based on the projection onto each transient state as follows:
\begin{align}
n^{\mathrm{B}}_c(t) &= \frac{a_L}{2\pi} \int^{\frac{2\pi}{a_L}}_0 dk \left| \left(
\vecb{u}^{\mathrm{B}}_{ck} \right)^{\dagger} \vecb{u}_{k}(t) \right|^2, \label{eq:pop-bloch} \\
n^{\mathrm{H}}_c(t) &= \frac{a_L}{2\pi} \int^{\frac{2\pi}{a_L}}_0 dk \left| \left(
\vecb{u}^{\mathrm{H}}_{ck}(t) \right)^{\dagger} \vecb{u}_{k}(t) \right|^2, \label{eq:pop-houston} \\
n^{\mathrm{PH}}_c(t) &= \frac{a_L}{2\pi} \int^{\frac{2\pi}{a_L}}_0 dk \left| \left(
\vecb{u}^{\mathrm{PH}}_{ck}(t) \right)^{\dagger} \vecb{u}_{k}(t) \right|^2. \label{eq:pop-pol-houston}
\end{align}

Hereafter, we examine the behavior of the population dynamics computed using Eqs.~(\ref{eq:pop-bloch}--\ref{eq:pop-pol-houston}) under several conditions in order to elucidate the nature of the populations associated with each basis.

\subsection{Population Dynamics Under Static Fields \label{subsec:tdse-quasi-static}}

\begin{figure}[htbp]
 \includegraphics[width=1.0\linewidth]{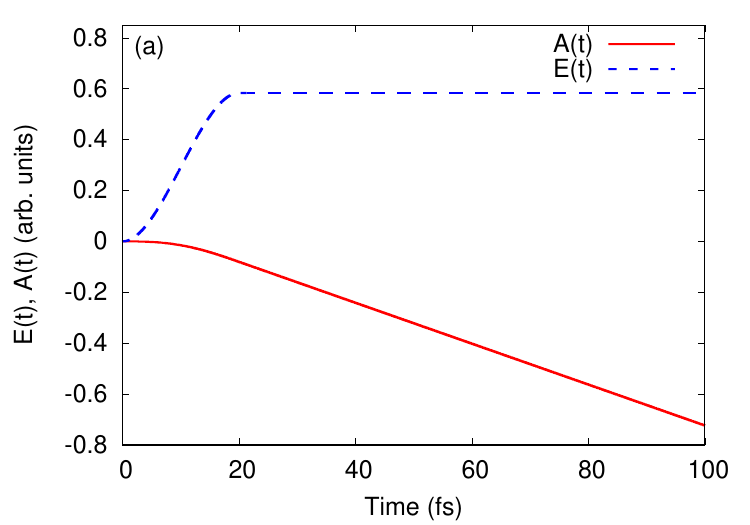}
 \includegraphics[width=1.0\linewidth]{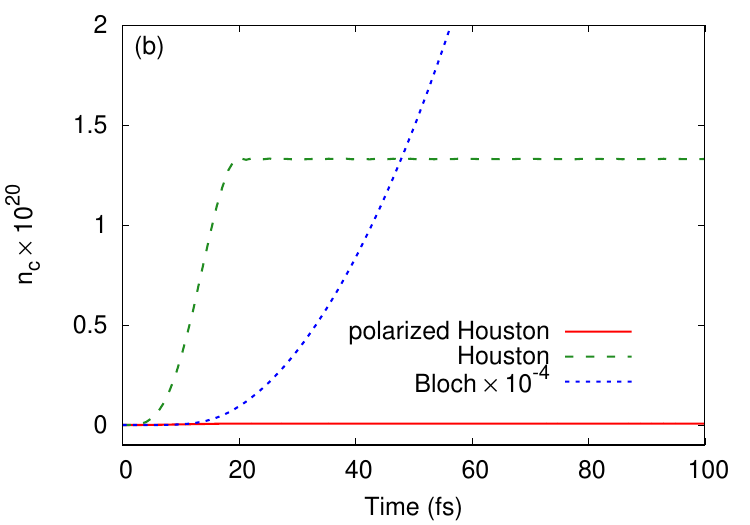}
\caption{\label{fig:static_fields} 
(a) The time profile of the applied fields given in Eq.~(\ref{eq:static-vector-potential}) and Eq.~(\ref{eq:static-electric-field}). (b) The computed conduction population dynamics with different methods: $n^{\mathrm{PH}}_c(t)$ in Eq.~(\ref{eq:pop-pol-houston}), $n^{\mathrm{H}}_c(t)$ in Eq.~(\ref{eq:pop-houston}), and $n^{\mathrm{B}}_c(t)$ in Eq.~(\ref{eq:pop-bloch}). The result, $n^{\mathrm{B}}_c(t)$, computed with the Bloch states is scaled by a factor of $10^{-4}$.
}
\end{figure}

We first look at the population dynamics under static fields. For this purpose, we employ the following form for the vector potential, $A(t)$ as shown in Fig.~\ref{fig:static_fields}~(a):
\begin{align}
A(t) = 
\begin{cases} 
  0  & (t < 0),  \\ 
  -E_{dc}T_{dc} \left[ \left( \frac{t}{T_{dc}} \right)^3 - \frac{1}{2} \left( \frac{t}{T_{dc}} \right)^4 \right] & (0 \leq t \leq T_{dc}), \\
  -E_{dc} \left( t - T_{dc} \right) - \frac{1}{2} E_{dc} T_{dc} & (T_{dc} < t).
\end{cases}
\label{eq:static-vector-potential}
\end{align}

\begin{widetext}
The corresponding electric field, $E(t)$, is given by  
\begin{align}
E(t) = -\frac{d}{dt} A(t) = 
\begin{cases} 
  0 & (t < 0), \\ 
  E_{dc} \left[ 3 \left( \frac{t}{T_{dc}} \right)^2 - 2 \left( \frac{t}{T_{dc}} \right)^3 \right] & (0 \leq t \leq T_{dc}), \\ 
  E_{dc} & (T_{dc} < t).
\end{cases}
\label{eq:static-electric-field}
\end{align}
Here, $E_{dc}$ represents the strength of the static field, and $T_{dc}$ denotes the rise time. In this work, we set $E_{dc}$ to $1$~V/m and $T_{dc}$ to $20$~fs.
\end{widetext}

Figure~\ref{fig:static_fields}~(a) displays the time profiles of the vector potential [Eq.~(\ref{eq:static-vector-potential})] and the corresponding electric field [Eq.~(\ref{eq:static-electric-field})]. While the vector potential increases continuously, the electric field reaches a constant value after the rise time, $T_{dc} = 20$~fs.

To evaluate the three formulations of dynamical population given in Eqs.~(\ref{eq:pop-bloch}--\ref{eq:pop-pol-houston}) under a static electric field, we compute the conduction band population using the field profile shown in Fig.~\ref{fig:static_fields}~(a). The resulting population dynamics, $n^{\mathrm{B}}_c(t)$, $n^{\mathrm{H}}_c(t)$, and $n^{\mathrm{PH}}_c(t)$, are shown in Fig.~\ref{fig:static_fields}~(b). The red solid line corresponds to the projection onto polarized Houston states, the green dashed line to Houston states, and the blue dotted line to Bloch states.

As seen in Fig.~\ref{fig:static_fields}, the conduction population $n^{\mathrm{B}}_c(t)$ computed with the Bloch states is significantly large and keeps increasing even after the applied electric field becomes constant ($t > T_{dc} $). Since the applied electric field is very weak ($E_0 = 1$~V/m), such significant generation of excited electrons appears inconsistent with the nature of insulators. This spurious excitation in the Bloch basis projection mainly originates from the fact that the intraband motion due to the wavevector shift, $\vecb{k} \rightarrow \vecb{k} + e\vecb{A}(t)/\hbar$, is not accounted for by the Bloch states, as the wavevector $\vecb{k}$ remains fixed in the Bloch states. These spurious excitations have been discussed in calculations using time-dependent density functional theory~\cite{PhysRevB.77.165104} and the quantum master equation~\cite{NJP_Sato_2019}. 

In contrast, the conduction population $n^{\mathrm{H}}_c(t)$ computed with the Houston states significantly suppresses spurious excitations, remaining constant after the applied field becomes constant ($t > T_{dc}$). Since projection onto the Houston states in the velocity gauge corresponds to projection onto the eigenstates of the field-free Hamiltonian in the length gauge, $n^{\mathrm{H}}_c(t)$ represents the population computed by projecting onto the bare bands without fields (see Appendix~\ref{appendix:sec:ex-hydrogen} and \ref{appendix:sec:ex-solids}). In real systems under a static field, electronic systems become polarized, modifying the wavefunction and energy levels. Consequently, the polarized system contains a certain population in the excited states of the field-free Hamiltonian. As the excitation associated with system polarization vanishes after the field application and the polarization disappears, this excitation is often interpreted as \textit{virtual} excitation~\cite{Sommer2016,McDonald_2017,PhysRevB.102.041125}. Hence, in Fig.~\ref{fig:static_fields}~(b), the population computed with the Houston states is dominated by virtual excitation. As a counterpart, \textit{real} excitation is often discussed as the population that remains after field irradiation. However, there is no clear definition of real versus virtual excitation under fields.

In contrast to $n^{\mathrm{B}}_c(t)$ and $n^{\mathrm{H}}_c(t)$, the conduction population $n^{\mathrm{PH}}_c(t)$ computed with the polarized Houston states is significantly suppressed throughout the investigated range in Fig.~\ref{fig:static_fields}~(b). This result indicates that the virtual excitation contribution to the temporal population is significantly suppressed when projected onto the polarized Houston states. This observation can be naturally understood by the fact that polarized Houston states, defined by diagonalizing the Hamiltonian with field-induced interband coupling terms, partially account for field-induced polarization effects. 

Although virtual population provides valuable insights into dynamic systems, a more intuitive population free from virtual excitation is often desired to interpret and analyze light-induced phenomena, such as the Pauli blocking effects in transient absorption spectra~\cite{doi:10.1126/science.1260311,Zurch2017,Schlaepfer2018}. In the following sections, we further examine the temporal population dynamics with different reference states: the Bloch, Houston, and polarized Houston states.

\subsection{Population Dynamics under Off-Resonant Laser Pulse \label{subsec:tdse-off-resonant-pulse}}

Here, we elucidate the population dynamics under a pulsed electric field in an off-resonant regime. For this purpose, we employ the following form for the vector potential:
\begin{align}
A(t) = -\frac{E_0}{\omega_0} \sin \left [\omega_0 \left (t - \frac{T_{\mathrm{pulse}}}{2} \right ) \right ] 
\cos^4 \left [\pi \left (\frac{t - \frac{T_{\mathrm{pulse}}}{2}}{T_{\mathrm{pulse}}} \right ) \right ]
\label{eq:pulse-vectorpotential}
\end{align}
in the domain $0 \leq t \leq T_{\mathrm{pulse}}$ and zero otherwise. Here, $E_0$ is the peak field strength, $\omega_0$ is the mean frequency, and $T_{\mathrm{pulse}}$ is the full pulse duration.

\begin{figure}[htbp]
 \includegraphics[width=1.0\linewidth]{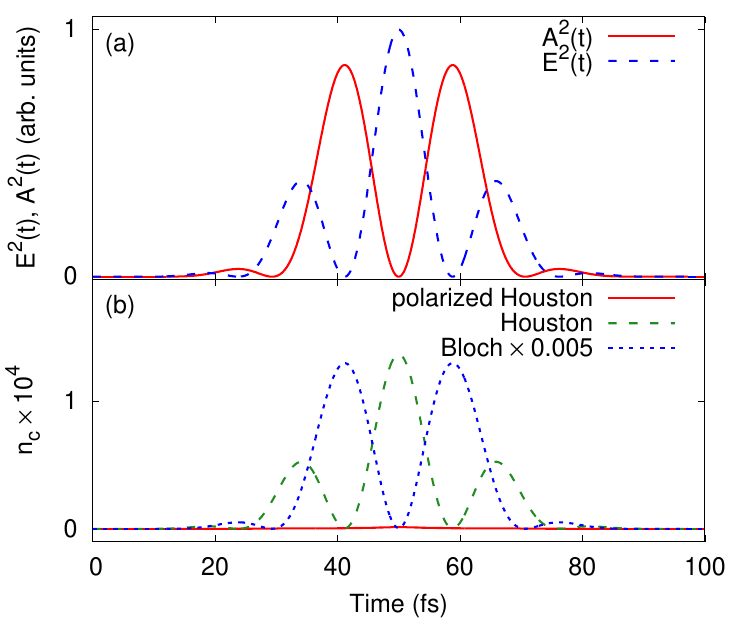}
\caption{\label{fig:pulse_off_resonant_weak} 
(a) The time profile of the applied fields given in Eq.~(\ref{eq:pulse-vectorpotential}). (b) The computed conduction population dynamics with different methods: $n^{\mathrm{PH}}_c(t)$ in Eq.~(\ref{eq:pop-pol-houston}), $n^{\mathrm{H}}_c(t)$ in Eq.~(\ref{eq:pop-houston}), and $n^{\mathrm{B}}_c(t)$ in Eq.~(\ref{eq:pop-bloch}). The result $n^{\mathrm{B}}_c(t)$, computed with the Bloch states, is scaled by a factor of $0.005$.
}
\end{figure}

To investigate the electron dynamics under off-resonant laser driving, we set $\omega_0$ to $0.1$~eV/$\hbar$ and $T_{\mathrm{pulse}}$ to $100$~fs. Figure~\ref{fig:pulse_off_resonant_weak}~(a) shows the time profiles of the square of the applied vector potential and the electric field. We compute the electron dynamics under these fields by solving Eq.~(\ref{eq:tdse-dimer-chain}) and evaluate the conduction population using Eqs.~(\ref{eq:pop-bloch}--\ref{eq:pop-pol-houston}).

 We first examine the dynamics in the weak field regime by setting $E_0$ to $1$~MV/cm in Eq.~(\ref{eq:pulse-vectorpotential}). Figure~\ref{fig:pulse_off_resonant_weak}~(b) shows the computed population dynamics. The conduction population, $n^{\mathrm{B}}_c(t)$, computed with the Bloch states, exhibits a much larger population compared to the other quantities, $n^{\mathrm{H}}_c(t)$ and $n^{\mathrm{PH}}_c(t)$. Its profile is closely aligned with the square of the applied vector potential, $A^2(t)$. As discussed in Sec.~\ref{subsec:tdse-quasi-static}, the conduction population computed with the Bloch states in Fig.~\ref{fig:pulse_off_resonant_weak}~(b) is dominated by spurious excitation due to the absence of intraband motion in the description using Bloch states. 

In contrast, the conduction population $n^{\mathrm{H}}_c(t)$, computed with the Houston states, as shown in Fig.~\ref{fig:pulse_off_resonant_weak}~(b), does not exhibit spurious conduction population since the contributions from intraband motion are accounted for in the Houston states. Instead, the time profile of $n^{\mathrm{H}}_c(t)$ closely follows the square of the electric field, $E^2(t)$, reflecting contributions from field-induced polarization and resulting virtual excitation.

Unlike $n^{\mathrm{B}}_c(t)$ and $n^{\mathrm{H}}_c(t)$, the spurious and virtual excitations in the conduction population $n^{\mathrm{PH}}_c(t)$, computed with polarized Houston states, are significantly suppressed. This is because the polarized Houston states accurately describe the intraband motion and polarization effects.

\begin{figure}[htbp]
 \includegraphics[width=1.0\linewidth]{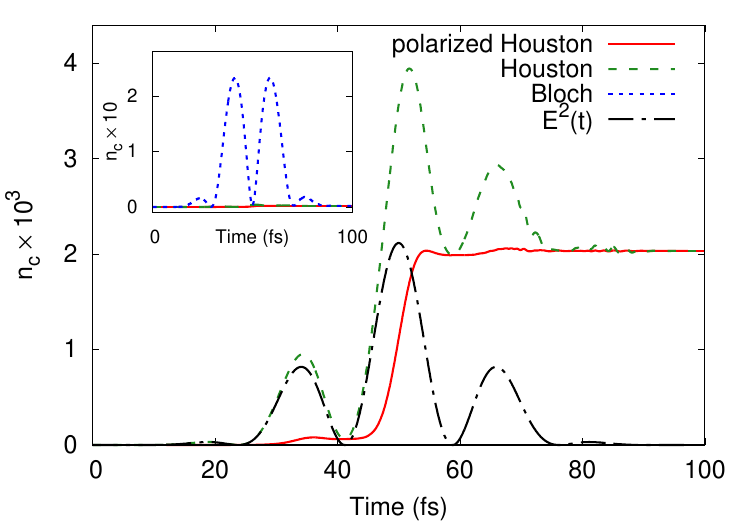}
\caption{\label{fig:pulse_off_resonant_strong} 
(a) The time profile of the applied fields given in Eq.~(\ref{eq:static-vector-potential}) and Eq.~(\ref{eq:static-electric-field}). (b) The computed conduction population dynamics with different methods: $n^{\mathrm{PH}}_c(t)$ in Eq.~(\ref{eq:pop-pol-houston}), $n^{\mathrm{H}}_c(t)$ in Eq.~(\ref{eq:pop-houston}), and $n^{\mathrm{B}}_c(t)$ in Eq.~(\ref{eq:pop-bloch}). The result $n^{\mathrm{B}}_c(t)$, computed with the Bloch states, is scaled by a factor of $10^{-4}$.
}
\end{figure}

Having demonstrated that the projection with polarized Houston states effectively suppresses spurious and virtual excitations in the evaluation of dynamical population analysis in the weak field regime, we next investigate the population dynamics in the strong field regime by setting $E_0$ to $4$~MV/m. Figure~\ref{fig:pulse_off_resonant_strong} presents the computed population dynamics using different equations, Eqs.~(\ref{eq:pop-bloch}--\ref{eq:pop-pol-houston}). As shown in the inset of Fig.~\ref{fig:pulse_off_resonant_strong}, the conduction population $n^{\mathrm{B}}_c(t)$, computed with the Bloch states, is significantly dominated by spurious excitation, making it difficult to extract relevant information about the field-induced dynamics. 

In contrast, the conduction population $n^{\mathrm{H}}_c(t)$, evaluated using Houston states, provides more insightful dynamics. In addition to the virtual excitation dynamics, whose time profile resembles the square of the applied electric field $E^2(t)$, one can observe real carrier injection dynamics via tunneling excitation, resulting in a residual population in the conduction band after the field.

Although the virtual excitation dynamics itself may provide valuable information, the virtual excitation contribution often complicates the investigation of real excitation dynamics, as the real excitation is partially obscured by the virtual carrier dynamics. As shown in Fig.~\ref{fig:pulse_off_resonant_strong}, the virtual excitation contributions are significantly suppressed in the conduction population $n^{\mathrm{PH}}_c(t)$, which is computed using polarized Houston states. Consequently, one can clearly observe significant carrier injection occurring only around the peak of the applied electric field, consistent with the physical picture of tunneling ionization~\cite{doi:10.1126/science.1260311}.

\subsection{Population Dynamics Under a Resonant Laser Pulse \label{subsec:tdse-resonant-pulse}}

\begin{figure}[htbp]
\includegraphics[width=1.0\linewidth]{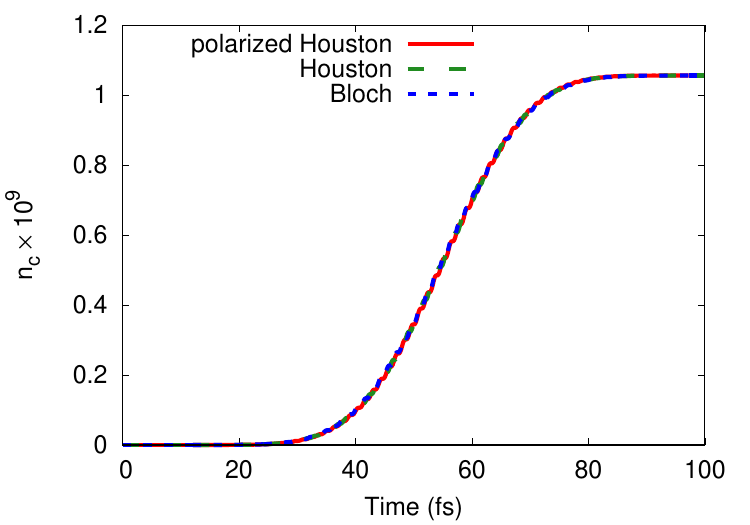}
\caption{\label{fig:pulse_resonant_weak} 
(a) The time profile of the applied fields, as given in Eq.~(\ref{eq:static-vector-potential}) and Eq.~(\ref{eq:static-electric-field}). (b) The computed conduction population dynamics using different methods: $n^{\mathrm{PH}}_c(t)$ in Eq.~(\ref{eq:pop-pol-houston}), $n^{\mathrm{H}}_c(t)$ in Eq.~(\ref{eq:pop-houston}), and $n^{\mathrm{B}}_c(t)$ in Eq.~(\ref{eq:pop-bloch}). The result $n^{\mathrm{B}}_c(t)$, computed with the Bloch states, is scaled by a factor of $10^{-4}$. 
}
\end{figure}

Here, we investigate the population dynamics under a resonant driving field by setting $E_0$ to $0.01$~MV/cm in Eq.~(\ref{eq:pulse-vectorpotential}) and $\omega_0$ to $1.55$~eV/$\hbar$, which is slightly larger than the gap $\Delta/\hbar=1.52$~eV/$\hbar$. Figure~\ref{fig:pulse_resonant_weak} shows the computed conduction population under the resonant driving field using different expressions, Eqs.~(\ref{eq:pop-bloch}--\ref{eq:pop-pol-houston}). In contrast to the responses observed under static and off-resonant driving fields discussed in previous sections, all three expressions produce similar population dynamics in the resonant regime, regardless of the presence of spurious or virtual excitations. This indicates that real photocarrier injection dominates the dynamics, rendering the contributions from spurious and virtual excitations negligible.

From another perspective, the field-induced intraband motion is largely suppressed in the resonant regime ($\omega \approx \Delta/\hbar$) compared to the deeply off-resonant regime ($\omega \ll \Delta/\hbar$), assuming a fixed driving field strength $E_0$. This suppression can be understood considering that the amplitude of the vector potential $A_0$ is proportional to $E_0/\omega$.

Based on the analysis of the static field, off-resonant driving field, and resonant driving field, we have demonstrated that, in contrast to the Bloch and Houston states, spurious and virtual excitations are significantly suppressed in the population computed using the polarized Houston states. This reflects the fact that polarized Houston states effectively capture the field-induced intraband motion and polarization contributions. Therefore, the projection onto polarized Houston states provides a natural choice for computing instantaneous band populations and real excitation dynamics under the influence of the fields.

\section{Reference States of the Relaxation Time Approximation in the Quantum Master Equation \label{sec:ref-state-q-master}}

In Sec. III, we examined the use of $u^{\mathrm{B}}_{b\vecb{k}}(\vecb{r})$, $u^{\mathrm{H}}_{b\vecb{k}}(\vecb{r},t)$, and $u^{\mathrm{PH}}_{b\vecb{k}}(\vecb{r},t)$ within the framework of the time-dependent Schr\"odinger equation, which governs the dynamics of closed quantum systems. However, our interest extends to open quantum systems, which are often of equal or greater relevance in realistic scenarios. To describe the dynamics of open systems, the quantum master equation~\cite{lidar2020lecturenotestheoryopen}, often combined with relaxation time approximations~\cite{PhysRevLett.73.902}, is widely employed. In the solid-state physics community, this formalism is commonly referred to as the semiconductor Bloch equations~\cite{PhysRevB.38.3342,PhysRevLett.73.902,Meier2007}. In this section, we explore the use of $u^{\mathrm{B}}_{b\vecb{k}}(\vecb{r})$, $u^{\mathrm{H}}_{b\vecb{k}}(\vecb{r},t)$, and $u^{\mathrm{PH}}_{b\vecb{k}}(\vecb{r},t)$ as reference states in the relaxation time approximation.

For practical analysis, we consider the light-induced electron dynamics described by the following quantum master equation:
\begin{align}
\frac{d}{dt}\rho_{\vecb{k}}(t) &= \frac{1}{i\hbar}\left [H_{\vecb{k} + e\vecb{A}(t)/\hbar}, \rho_{\vecb{k}}(t) \right ] 
+ \hat{D}\left [\rho_{\vecb{k}}(t) \right ],
\label{eq:q-master}
\end{align}
where $\rho_{\vecb{k}}(t)$ is the one-body reduced density matrix associated with the initial Bloch wavevector $\vecb{k}$, and $\hat{D}\left [\rho_{\vecb{k}}(t) \right ]$ is the relaxation operator. Note that if the relaxation operator is omitted, Eq.~(\ref{eq:q-master}) reduces to Eq.~(\ref{eq:tdse}).

\begin{widetext}
In this work, we employ the following relaxation time approximation for the relaxation operator:
\begin{align}
\hat{D}\left [\rho_{\vecb{k}}(t) \right ] = 
& -\frac{1}{T_1}
\sum_{b} \big|u^{\mathrm{Ref}}_{b\vecb{k}}(t)\big\rangle 
\Big[\big\langle u^{\mathrm{Ref}}_{b\vecb{k}}(t) \big| \rho_{\vecb{k}}(t) \big|u^{\mathrm{Ref}}_{b\vecb{k}}(t)\big\rangle - \rho^{\mathrm{FD}}\left(\epsilon^{\mathrm{Ref}}_{b\vecb{k}}(t)\right)\Big]
\big\langle u^{\mathrm{Ref}}_{b\vecb{k}}(t)\big| \nonumber \\
& - \frac{1}{T_2}
\sum_{b \neq b'} \big|u^{\mathrm{Ref}}_{b\vecb{k}}(t)\big\rangle 
\big\langle u^{\mathrm{Ref}}_{b\vecb{k}}(t) \big| \rho_{\vecb{k}}(t) \big|u^{\mathrm{Ref}}_{b'\vecb{k}}(t)\big\rangle 
\big\langle u^{\mathrm{Ref}}_{b'\vecb{k}}(t)\big|,
\label{eq:relaxation-time-approx}
\end{align}
where $T_1$ is the longitudinal relaxation time associated with population relaxation, while $T_2$ is the transverse relaxation time associated with decoherence.
\end{widetext}

In Eq.~(\ref{eq:relaxation-time-approx}), we introduced the reference states $\big|u^{\mathrm{Ref}}_{b\vecb{k}}(t)\big\rangle$ and the corresponding single-particle energies $\epsilon^{\mathrm{Ref}}_{b\vecb{k}}(t)$ to describe the proper relaxation to the Fermi–Dirac distribution, $\rho^{\mathrm{FD}}$, given by:
\begin{align}
\rho^{\mathrm{FD}}(\epsilon) = \frac{1}{e^{(\epsilon - \mu)/k_{\mathcal{B}}T_e} + 1},
\end{align}
where $\epsilon$ is the single-particle energy, $\mu$ is the chemical potential, and $T_e$ is the electron temperature.

As reference states $\big|u^{\mathrm{Ref}}_{b\vecb{k}}(t)\big\rangle$ and their corresponding single-particle energies $\epsilon^{\mathrm{Ref}}_{b\vecb{k}}(t)$, we employ the following combinations: The first set is consist of the Bloch states $u_{b\vecb{k}}(\vecb{x})$ and the single-particle energies $\epsilon_{b\vecb{k}}$, as defined in Eq.~(\ref{eq:bloch-basis}). Another set consists of the Houston states $u^{\mathrm{H}}_{b\vecb{k}}(\vecb{r},t)$ and the instantaneous eigenenergies $\epsilon_{b,\vecb{k} + e\vecb{A}(t)/\hbar}$ of the instantaneous Hamiltonian $\hat{h}_{\vecb{k} + e\vecb{A}(t)/\hbar}$. The third set is consist of the polarized Houston states $u^{\mathrm{PH}}_{b\vecb{k}}(\vecb{r},t)$ and the instantaneous eigenenergies $\epsilon^{P}_{b\vecb{k}}(t)$ of the effective Hamiltonian in Eq.~(\ref{eq:effective-ham-in-length-gauge}).

We note that when the Houston states $u^{\mathrm{H}}{b\vecb{k}}(\vecb{r},t)$ are used as the reference states $\big|u^{\mathrm{Ref}}{b\vecb{k}}(t)\big\rangle$, the resulting relaxation operator is identical to the one used in previous works~\cite{PhysRevB.99.224301,PhysRevB.99.214302,NJP_Sato_2019}. Furthermore, as discussed in Appendix~\ref{appendix:sec:houston-SBE}, the resulting equation of motion is identical to those of the semiconductor Bloch equations~\cite{PhysRevB.38.3342,PhysRevLett.73.902,Meier2007}.

To evaluate the choice of reference states, we investigate the field-induced electron dynamics using the one-dimensional dimer-chain model Hamiltonian defined in Eq.~(\ref{eq:ham-dimer-chain}). For the relaxation operator in Eq.~(\ref{eq:relaxation-time-approx}), we set $T_1 = T_2 = 20$~fs. For the Fermi–Dirac distribution, both $\mu$ and $T_e$ are set to zero.

We compute the electron dynamics under the time-dependent vector potential $A(t)$ defined in Eq.~(\ref{eq:static-vector-potential}), which corresponds to the static electric field after the ramping time $T_{\mathrm{dc}}$, as shown in Fig.~\ref{fig:static_fields}(a). As an observable, we evaluate the current $J(t)$ induced by the fields using the following expressions:
\begin{align}
\hat{J}_{k}(t) &= -\frac{e}{\hbar} \frac{\partial H_{k}}{\partial k} \bigg|_{k = k + eA(t)/\hbar}, \nonumber \\
J(t) &= \frac{a_L}{2\pi}\int_{\vecb{k}} \mathrm{Tr}\left[\hat{J}_{k}(t) \rho_{k}(t)\right].
\end{align}

\begin{figure}[htbp]
\includegraphics[width=1.0\linewidth]{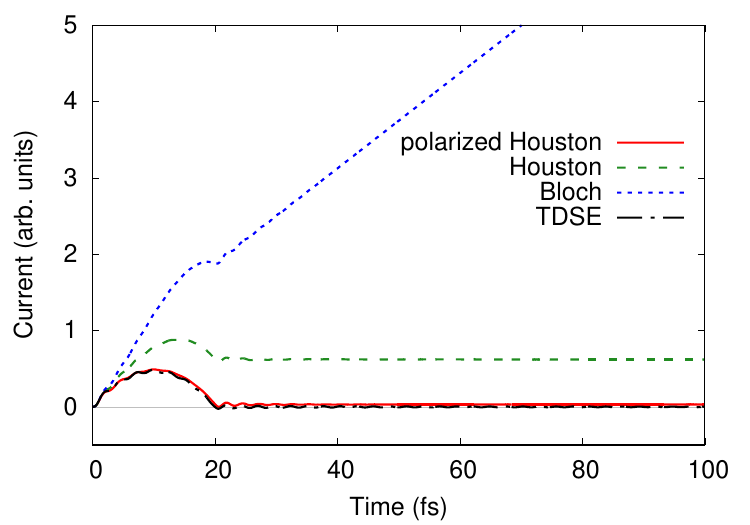}
\caption{\label{fig:q_master_static} 
Time profile of the computed current under the fields defined in Eq.~(\ref{eq:static-vector-potential}) with the quantum master equation. Different reference states in the relaxation time approximation are employed: the polarized Houston states, the Houston states, and the Bloch states. For reference, the result with the time-dependent Schr\"odinger equation (TDSE) is also shown.}
\end{figure}

Figure~\ref{fig:q_master_static} shows the computed current using the quantum master equation with different reference states. The result with the time-dependent Schr\"odinger equation, without the relaxation operator, is also shown as the black dash-dot line. Although the current should be zero in insulators under a weak static field, the result computed with the Bloch state (blue dotted line) continuously increases even after the applied field becomes constant ($t > T_{dc}$). Thus, the relaxation time approximation with the Bloch state may induce a large unphysical current, obscuring the physical current~\cite{NJP_Sato_2019}. This unphysical current arises from the lack of intraband motion in the description of the Bloch states. The current computed with the Houston states (green dashed line) results in a non-zero constant current after the applied field becomes constant. This unphysical current stems from the absence of field-induced polarization effects in the Houston states. Recently, Terada \textit{et al.} reported this unphysical current in the relaxation time approximation and proposed a method to remove it by considering field-induced effects in the coupling with the bath.~\cite{PhysRevB.109.L180302}. 

The current computed with the polarized Houston states (red solid line) significantly suppresses the current after the applied field becomes constant and is very similar to the result of the time-dependent Schr\"odinger equation without any relaxation effects. This result indicates that the polarized Houston states incorporate field-induced polarization effects and intraband motion, and thus, relaxation with respect to the polarized Houston states suppresses spurious excitations. Furthermore, one may analytically prove that the polarized Houston states are the exact solutions of the time-dependent Schr\"odinger equation in the static field and linear response limits. As a result, the current computed using the relaxation time approximation with the polarized Houston states becomes exactly zero for insulators in the static field and linear response limits, eliminating the spurious current associated with the Houston states (see Appendix~\ref{appendix:sec:adiabatic-evolution}).

\section{Summary}

In this work, we introduced \textit{polarized Houston states} as an extension of conventional Houston states to investigate light-induced electron dynamics in solids. Unlike conventional Houston states, which only account for intraband motion under external fields, polarized Houston states incorporate field-induced interband transitions, naturally capturing polarization effects such as the Stark shift. This refined basis allows for a more accurate representation of carrier dynamics by significantly suppressing virtual excitations, which often obscure real population dynamics in driven quantum systems.

To assess the effectiveness of polarized Houston states, we systematically analyzed the dynamical evolution of band populations in a one-dimensional dimer-chain model, comparing projections onto static Bloch states, conventional Houston states, and polarized Houston states. Our results show that projections onto static Bloch states suffer from significant spurious excitations due to their neglect of intraband motion, while conventional Houston states introduce persistent virtual excitations during field irradiation. In contrast, projections onto polarized Houston states effectively eliminate both spurious and virtual excitations, yielding a physically meaningful description of electron dynamics.

In the strong-field regime, where real carrier injection becomes prominent, population analysis using polarized Houston states reveals a sharp increase in conduction band occupation near the peak of the electric field, consistent with tunneling ionization mechanisms. This signature is largely obscured when using conventional Houston or Bloch states, where virtual and spurious excitations dominate the population response. Our findings demonstrate that polarized Houston states offer a significantly improved framework for analyzing ultrafast laser-driven excitation processes in semiconductors and insulators, especially in strong-field regimes where distinguishing real from virtual excitations is crucial.

Beyond population dynamics, we applied polarized Houston states within the relaxation-time approximation of the quantum master equation. Previous studies have reported that using conventional Houston states in this context leads to spurious DC currents in insulators under static fields, due to the neglect of field-induced polarization effects~\cite{PhysRevB.109.L180302}. We show that incorporating polarized Houston states into the relaxation-time approximation completely eliminates unphysical DC responses, providing a more accurate description of relaxation and dissipation processes in driven quantum systems.

These results establish polarized Houston states as a robust and physically motivated framework for analyzing ultrafast electron dynamics in solids. By enabling a more accurate extraction of real carrier injection dynamics and providing a refined foundation for modeling relaxation phenomena, polarized Houston states improve the reliability of theoretical descriptions of field-driven quantum transport. Their ability to naturally account for field-induced modifications in electronic structure would provide a more straightforward understanding of light-matter interactions in condensed matter physics.

\begin{acknowledgments}
This work was supported by JSPS KAKENHI Grant Numbers JP21H01842 and JP25K08501, MEXT Promotion of Development of a Joint Usage/Research System Project: Coalition of Universities for Research Excellence Program (CURE) Grant Number JPMXP1323015474, the International Collaborative Research Program of Institute for Chemical Research, Kyoto University (Grant No. 2024-19), and the Research Foundation for Opto-Science and Technology. We also acknowledge support from the Marie Sklodowska-Curie Doctoral Network TIMES, grant No. 101118915, and SPARKLE grant No. 101169225; the Italian Ministry of University and Research (MUR) under the PRIN 2022 Grant No 2022PX279E\_003; Next Generation EU Partenariato Esteso NQSTI-Spoke 2 (THENCE-PE00000023). This work used computational resources of the HPC systems at the Max Planck Computing and Data Facility (MPCDF), and the Supercomputer Center, the Institute for Solid State Physics, the University of Tokyo.

\end{acknowledgments}

\appendix

\section{Instantaneous ground state population of isolated systems \label{appendix:sec:ex-hydrogen}}

To obtain a clear picture of the instantaneous population of a state under the influence of gauge fields, we consider the adiabatic electron dynamics in a hydrogen atom. We use the time-dependent Schr\"odinger equation in the length gauge, expressed as:
\begin{align}
i\hbar \frac{\partial}{\partial t}\psi(\vecb{r},t) = \left[\frac{\vecb{p}^2}{2m} - \frac{e^2}{4\pi \epsilon_0} \frac{1}{r} + e\vecb{E}(t)\cdot \vecb{r}\right] \psi(\vecb{r},t),
\end{align}
where $\vecb{E}(t)$ is the homogeneous electric field applied to the system. As the initial condition, we set the wavefunction to the hydrogen ground state as $\psi(\vecb{r}, t=-\infty) = \phi_{1s}(\vecb{r})$. Furthermore, we assume the applied electric field takes the following form:
\begin{align}
\vecb{E}(t) = -\vecb{A}_0 \frac{1}{\sqrt{2\pi \sigma^2}} e^{-\frac{t^2}{2\sigma^2}},
\label{appendix:eq:e-field}
\end{align}
where $\vecb{A}_0$ is a constant vector, and $\sigma$ represents the duration of the applied field.

Since the electric field approaches zero in the limit $(\sigma \rightarrow \infty)$, the time-dependent wavefunction becomes identical to the hydrogen ground state, except for a phase factor: $\psi(\vecb{r},t) = e^{i\phi(t)}\phi_{1s}(\vecb{r})$. Thus, the probability of finding the system in the ground state remains unity, as $|\langle \phi_{1s}|\psi(t)\rangle|^2 = 1$.

We also analyze the same phenomenon using the velocity gauge. For this purpose, we apply a gauge transformation to the wavefunction,
\begin{align}
\tilde{\psi}(\vecb{r},t) = \exp\left[-\frac{i}{\hbar} e\vecb{A}(t)\cdot\vecb{r}\right]\psi(\vecb{r},t),
\end{align}
where the vector potential is given by $\vecb{A}(t) = -\int_{-\infty}^{t} dt' \vecb{E}(t')$. The transformed wavefunction, $\tilde{\psi}(\vecb{r},t)$, satisfies the following time-dependent Schr\"odinger equation in the velocity gauge:
\begin{align}
i\hbar \frac{\partial}{\partial t} \tilde{\psi}(\vecb{r},t) = \left[\frac{\left(\vecb{p} + e\vecb{A}(t)\right)^2}{2m} - \frac{e^2}{4\pi \epsilon_0} \frac{1}{r}\right] \tilde{\psi}(\vecb{r},t).
\label{appendix:eq:tdse-hydrogen-vel}
\end{align}
Imposing the vector potential is zero at $t=-\infty$, the initial condition for $\tilde{\psi}(\vecb{r},t)$ is given by $\tilde{\psi}(\vecb{r}, t=-\infty) = \psi(\vecb{r}, t=-\infty) = \phi_{1s}(\vecb{r})$.

By employing the adiabaticity theorem, the time-dependent wavefunction $\tilde \psi(\vecb r,t)$ is an eigenstate of the instantaneous Hamiltoinan and is proportional to $\exp \left [-\frac{i}{\hbar}e \vecb A(t) \cdot \vecb r \right ] \phi_{1s}(\vecb r)$ in the limit of ($\sigma \rightarrow \infty$). Hence, a naive overlap between $|\tilde \psi(t)\rangle$ and $|\phi_{1s}\rangle$ does not yield unity as
\begin{align}
|\langle \tilde \psi (t)|\phi_{1s}\rangle|^2 = |\langle \phi_{1s}| e^{i \vecb A(t) \cdot \vecb r}|\phi_{1s}\rangle|^2 \neq |\langle \phi_{1s}|\phi_{1s}\rangle|^2 = 1.
\label{appendix:eq:naive-overlap-velocity-gauge}
\end{align}

Since the physical dynamics must be identical between the length and velocity gauges, the hydrogen atom should not be excited by the laser field given by Eq.~(\ref{appendix:eq:e-field}) in the limit ($\sigma \rightarrow \infty$). Therefore, Eq.~(\ref{appendix:eq:naive-overlap-velocity-gauge}) indicates that a naive overlap between the propagated state $|\tilde \psi(t)\rangle$ and the unperturbed state $|\phi_{1s}\rangle$ does not correspond to the probability of remaining in the state $|\phi_{1s}\rangle$ in the velocity gauge, due to the presence of the gauge field $\vecb A(t)$.

One may evaluate the correct overlap between $|\tilde \psi(t)\rangle$ and the unperturbed states by transforming the wavefunction to its length gauge representation as $|\psi(t)\rangle = \exp \left [\frac{i}{\hbar}e \vecb A(t) \cdot \vecb r \right ]|\tilde \psi(t)\rangle$ and then evaluating the overlap in the length gauge as $\langle \psi(t)|\phi_{1s}\rangle$. Another equivalent method to compute the correct overlap is to first provide the instantaneous eigenstates of the time-dependent Hamiltonian and then take the overlap between the time-dependent wavefunction $|\tilde \psi(t)\rangle$ and those instantaneous eigenstates in the velocity gauge. For example, the instantaneous ground state of the Hamiltonian in Eq.~(\ref{appendix:eq:tdse-hydrogen-vel}) is given by
\begin{align}
|\tilde \phi_{1s}(t)\rangle = \exp \left [-\frac{i}{\hbar}e \vecb A(t)\cdot \vecb r \right ]|\phi_{1s}\rangle.
\end{align}
By taking the overlap between $|\tilde \psi(t)\rangle$ and $|\tilde \phi_{1s}(t)\rangle$, the evaluated overlap becomes identical to that evaluated in the length gauge as $|\langle \tilde \psi(t)|\tilde \phi_{1s}(t)\rangle|^2 = |\langle \psi(t)|\phi_{1s}\rangle|^2$. For isolated systems, the two approaches to computing the correct overlap are identical. However, as will be shown later, the instantaneous eigenstate approach is more suitable for periodic systems.

\section{Instantaneous population of periodic systems \label{appendix:sec:ex-solids}}

Based on the discussion in the previous section, we discuss the population of a band in solid-state systems in the presence of a gauge field. For this purpose, we consider an $N$-electron system described by the following many-body Hamiltonian:
\begin{align}
\hat H_0 = \sum^N_{j=1}\left [\frac{\vecb p^2_j}{2m} + v(\vecb r_j)\right ],
\label{appendix:eq:ham-gs-many-elec}
\end{align}
where $v(\vecb r)$ is a one-body potential. We assume a spatial periodicity for the one-body potential as $v(\vecb r + \vecb a)=v(\vecb r)$ with lattice vectors $\vecb a$. Here, we ignore the explicit electron-electron interaction in the Hamiltonian, assuming that the effective mean-field potential $v(\vecb r)$ accounts for the many-body effects. The eigenstates of the Hamiltonian, Eq.~(\ref{appendix:eq:ham-gs-many-elec}), satisfy the following time-independent Schr\"odinger equation:
\begin{align}
\hat H_0 \Phi_n(\vecb x_1, \cdots, \vecb x_N) = E_n \Phi_n(\vecb x_1, \cdots, \vecb x_N),
\end{align}
where $\vecb x_j$ represents a set of real-space and spin coordinates as $\vecb x_j = (\vecb r_j, \sigma_j)$. Since the many-body Hamiltonian contains only the one-body parts, the eigenstates $\Phi_n(\vecb x_1, \cdots, \vecb x_N)$ can be described by a Slater determinant consisting of single-particle orbitals $\phi_k (\vecb x)$. Each single-particle orbital satisfies the following one-body Schr\"odinger equation:
\begin{align}
\left [\frac{\vecb p^2}{2m} + v(\vecb r)\right ]\phi_k(\vecb x) = \epsilon_k \phi_k(\vecb x).
\end{align}

Since the one-body potential has spatial periodicity, the Bloch theorem can be applied with the Born--von Karman boundary condition. As a result, the single-particle orbitals can be written in the following form:
\begin{align}
\phi_k(\vecb x) = e^{i\vecb k \cdot \vecb r}u_{b\vecb k}(\vecb x),
\end{align}
where $\vecb k$ is the Bloch wavevector, $b$ is the band index, and $u_{b\vecb k}(\vecb x)$ is the Bloch wavefunction. Note that the Bloch wavefunction has the same spatial periodicity as $v(\vecb r)$. The periodic part of the Bloch state satisfies the following equation:
\begin{align}
\left [\frac{\left (\vecb p + \hbar \vecb k \right )^2}{2m} + v(\vecb r)\right ]
u_{b\vecb k}(\vecb x) = \epsilon_{b\vecb k}
u_{b\vecb k}(\vecb x).
\label{appendix:eq:static-bloch-state}
\end{align}

\begin{widetext}
We then consider the dynamics of the $N$-electron system under a homogeneous electric field. The time evolution of the many-body wavefunction can be described by the following time-dependent Schr\"odinger equation:
\begin{align}
i\hbar \frac{\partial}{\partial t} \Psi(\vecb{x}_1, \cdots, \vecb{x}_N, t) = \left [\hat{H}_0 + \sum_{j=1}^N e \vecb{r}_j \cdot \vecb{E}(t) \right ] \Psi(\vecb{x}_1, \cdots, \vecb{x}_N, t).
\end{align}

\end{widetext}
Similarly to the time-independent case, the many-body wavefunction $\Psi(\vecb{x}_1, \cdots, \vecb{x}_N, t)$ can be described by a single Slater determinant, and each single-particle orbital satisfies the following one-body Schrödinger equation:
\begin{align}
i\hbar \frac{\partial}{\partial t} \psi_k(\vecb{x}, t) = \left [\frac{\vecb{p}^2}{2m} + v(\vecb{r}) + e \vecb{r} \cdot \vecb{E}(t) \right ] \psi_k(\vecb{x}, t).
\end{align}

Next, we consider the following gauge transformation to the velocity gauge:
\begin{align}
\psi_k(\vecb{x}, t) = \exp \left [ \frac{i}{\hbar} e \vecb{A}(t) \cdot \vecb{r} \right ] \tilde{\psi}_k(\vecb{x}, t).
\end{align}
In this gauge, the wavefunction satisfies the following Schr\"odinger equation:
\begin{align}
i\hbar \frac{\partial}{\partial t} \tilde{\psi}_k(\vecb{x}, t) = \left [\frac{\left (\vecb{p} + e \vecb{A}(t) \right )^2}{2m} + v(\vecb{r})\right ] \tilde{\psi}_k(\vecb{x}, t).
\label{appendix:eq:tdse-sp-solids-velocity}
\end{align}
Since the Hamiltonian in Eq.~(\ref{appendix:eq:tdse-sp-solids-velocity}) has the same periodicity as the potential $v(\vecb{r})$, one may apply the Born–von Karman boundary condition to $\tilde{\psi}_k(\vecb{x}, t)$ and employ the Bloch theorem. As a result, the single-particle orbitals can be written in the following form:
\begin{align}
\tilde{\psi}_k(\vecb{x}, t) = e^{i\vecb{k} \cdot \vecb{r}} u_{b\vecb{k}}(\vecb{x}, t).
\end{align}
Furthermore, the periodic part of the time-dependent Bloch states satisfy the following equation of motion:
\begin{align}
i\hbar \frac{\partial}{\partial t} u_{b \vecb k}(\vecb{x}, t) = \left [\frac{\left (\vecb{p} + \hbar \vecb k + e \vecb{A}(t) \right )^2}{2m} + v(\vecb{r})\right ] u_{b \vecb k}(\vecb{x}, t).
\label{eq:tdse-sp-solids-velocity-bloch}
\end{align}

We then consider the probability of finding the many-body system in the ground state. According to the discussion in Appendix~\ref{appendix:sec:ex-hydrogen}, one may consider to evaluate the overlap between the time-dependent many-body state $\Psi(\vecb{x}_1, \cdots, \vecb{x}_N, t)$ and the ground state $\Phi_0(\vecb{x}_1, \cdots, \vecb{x}_N)$ in the length gauge by transforming $\tilde{\psi}_k(\vecb{x},t)$ to $\psi_k(\vecb{x}, t)$. However, the naive gauge transformation is not suitable for this purpose because the time-dependent wavefunction $\psi(\vecb{x},t)$ in the length gauge may not satisfy the Born–von Karman boundary condition due to the phase factor $\exp \left [\frac{i}{\hbar}e \vecb{A}(t) \cdot \vecb{r}\right ]$. This issue arises because the boundary condition applies to the velocity gauge wavefunction $\tilde{\psi}(\vecb{x},t)$.

\begin{widetext}
As discussed in Appendix~\ref{appendix:sec:ex-hydrogen}, another way to compute the correct overlap is by using an instantaneous eigenstate of the Hamiltonian in the velocity gauge. The instantaneous many-body eigenstate is given by the following Schr\"odinger equation:
\begin{align}
\sum_{j}^{N}\left [\frac{\left( \vecb{p}_j + e\vecb{A}(t) \right )^2}{2m} + v(\vecb{r}_j)\right ] \tilde{\Phi}_{n}(\vecb{x}_1, \cdots, \vecb{x}_N, t) = E_n \tilde{\Phi}_{n}(\vecb{x}_1, \cdots, \vecb{x}_N, t).
\end{align}

The many-body wavefunction $\tilde{\Phi}_n(\vecb{x}_1, \cdots, \vecb{x}_N, t)$ can be described by a Slater determinant consisting of single-particle orbitals $\tilde{\phi}_k(\vecb{x}, t)$ that satisfy the following equation:
\begin{align}
\left [\frac{\left( \vecb{p} + e\vecb{A}(t) \right )^2}{2m} + v(\vecb{r}) \right ] \tilde{\phi}_k (\vecb{x}, t)
= \epsilon_{k} \tilde{\phi}_k (\vecb{x}, t).
\label{appendix:eq:instantaneous-eigenstates-velocity-gauge-solids}
\end{align}

By applying the Bloch theorem, we have
\begin{align}
\tilde{\phi}_k(\vecb{x}, t) = e^{i\vecb{k} \cdot \vecb{r}} u_{b,\vecb{k} + e\vecb{A}(t)/\hbar}(\vecb{x}),
\end{align}
where the periodic part of the Bloch state, $u_{b,\vecb{k} + e\vecb{A}(t)/\hbar}(\vecb{x})$, satisfies the following equation:
\begin{align}
\left [\frac{\left (\vecb{p} + \hbar \vecb{k} + e\vecb{A}(t) \right )^2}{2m} + v(\vecb{r})\right ]
u_{b,\vecb{k} + e\vecb{A}(t)/\hbar}(\vecb{x})
= \epsilon_{b,\vecb{k} + e\vecb{A}(t)/\hbar}
u_{b,\vecb{k} + e\vecb{A}(t)/\hbar}(\vecb{x}).
\end{align}
Note that $u_{b,\vecb{k} + e\vecb{A}(t)/\hbar}(\vecb{x})$ is simply the momentum-shifted ($\vecb{k} \rightarrow \vecb{k} + e\vecb{A}(t)/\hbar$) Bloch orbital from Eq.~(\ref{appendix:eq:static-bloch-state}).

Since both the time-dependent many-body wavefunction $\tilde{\Psi}(\vecb{x}_1, \cdots, \vecb{x}_N, t)$ and the many-body instantaneous eigenstate $\tilde{\Phi}_n(\vecb{x}_1, \cdots, \vecb{x}_N, t)$ satisfy the same Born--von Karman boundary condition, one may naturally consider the overlap as
\begin{align}
\left | \langle \tilde{\Phi}_n (t)| \tilde{\Psi}(t)\rangle  \right |^2 
= \left | 
\int d\vecb{x}_1 \cdots d\vecb{x}_N \tilde{\Phi}^*_n(\vecb{x}_1, \cdots, \vecb{x}_N, t)
\tilde{\Psi}(\vecb{x}_1, \cdots, \vecb{x}_N, t)
\right |^2.
\label{appendix:eq:natural-overlap-td-gs-velocity}
\end{align}

As an example of a practical evaluation of the probability, we consider a semiconductor with a single valence band, where only a single band $b=v$ is occupied in the ground state, and the other bands are empty. The ground state of such a system can be described by the following Slater determinant:
\begin{align}
\Phi_0(\vecb x_1, \cdots, \vecb x_N) = \frac{1}{\sqrt{N!}}
\begin{vmatrix}
e^{i\vecb k_1 \cdot \vecb r_1}u_{v \vecb k_1}(\vecb x_1) & \cdots &  e^{i\vecb k_N \cdot \vecb r_1}u_{v \vecb k_N}(\vecb x_1) \\
\vdots & \ddots & \vdots \\
e^{i\vecb k_1 \cdot \vecb r_N}u_{v \vecb k_1}(\vecb x_N) & \cdots &  e^{i\vecb k_N \cdot \vecb r_N}u_{v \vecb k_N}(\vecb x_N) \\
\end{vmatrix}.
\end{align}

Likewise, the time-dependent many-body wavefunction and the instantaneous ground state can be described as
\begin{align}
\tilde \Psi_0(\vecb x_1, \cdots, \vecb x_N, t) = \frac{1}{\sqrt{N!}}
\begin{vmatrix}
e^{i\vecb k_1 \cdot \vecb r_1}u_{\vecb k_1}(\vecb x_1, t) & \cdots &  e^{i\vecb k_N \cdot \vecb r_1}u_{\vecb k_N}(\vecb x_1, t) \\
\vdots & \ddots & \vdots \\
e^{i\vecb k_1 \cdot \vecb r_N}u_{\vecb k_1}(\vecb x_N, t) & \cdots &  e^{i\vecb k_N \cdot \vecb r_N}u_{\vecb k_N}(\vecb x_N, t) \\
\end{vmatrix},
\end{align}
and
\begin{align}
\tilde \Phi_0(\vecb x_1, \cdots, \vecb x_N, t) = \frac{1}{\sqrt{N!}}
\begin{vmatrix}
e^{i\vecb k_1 \cdot \vecb r_1}u_{v, \vecb k_1 + e\vecb A(t)/\hbar}(\vecb x_1) & \cdots &  e^{i\vecb k_N \cdot \vecb r_1}u_{v, \vecb k_N + e\vecb A(t)/\hbar}(\vecb x_1) \\
\vdots & \ddots & \vdots \\
e^{i\vecb k_1 \cdot \vecb r_N}u_{v, \vecb k_1 + e\vecb A(t)/\hbar}(\vecb x_N) & \cdots &  e^{i\vecb k_N \cdot \vecb r_N}u_{v, \vecb k_N + e\vecb A(t)/\hbar}(\vecb x_N) \\
\end{vmatrix}.
\label{appendix:eq:inst-gs-slater-solids}
\end{align}

Hence, the overlap in Eq.~(\ref{appendix:eq:natural-overlap-td-gs-velocity}) for the single-band semiconductor is evaluated as
\begin{align}
\left | \langle \tilde \Phi_0 (t) | \tilde \Psi(t) \rangle \right |^2 
& = \left | \int d\vecb x \, u^*_{v, \vecb{k}_1 + e \vecb{A}(t)/\hbar}(\vecb{x}) \, u_{\vecb{k}_1}(\vecb{x}, t) \right |^2 
\times \left | \int d\vecb x \, u^*_{v, \vecb{k}_2 + e \vecb{A}(t)/\hbar}(\vecb{x}) \, u_{\vecb{k}_2}(\vecb{x}, t) \right |^2 \nonumber \\
& \times \cdots \times \left | \int d\vecb x \, u^*_{v, \vecb{k}_N + e \vecb{A}(t)/\hbar}(\vecb{x}) \, u_{\vecb{k}_N}(\vecb{x}, t) \right |^2.
\label{appendix:eq:gs-occupation-solids}
\end{align}

\end{widetext}
The overlap of the many-body wavefunctions is naturally evaluated as the product of the overlaps between the time-dependent Bloch function $u_{\vecb{k}}(\vecb{x}, t)$ and the momentum-shifted Bloch function $u_{v, \vecb{k} + e \vecb{A}(t)/\hbar}(\vecb{x})$. Note that the momentum-shifted Bloch function is simply the instantaneous eigenstate of the time-dependent Hamiltonian in Eq.~(\ref{appendix:eq:instantaneous-eigenstates-velocity-gauge-solids}), commonly known as a Houston state~\cite{PhysRev.57.184}.

The population of the ground-state, or equivalently the zero-particle zero-hole state, is evaluated by Eq.~(\ref{appendix:eq:gs-occupation-solids}). Similarly, one may evaluate the population of a one-particle one-hole state as
\begin{align}
n_{vc, \vecb{k}_1 + e \vecb{A}(t)/\hbar} & =
\left | \langle \tilde \Phi_0(t) | \hat{a}^{\dagger}_{v, \vecb{k}_1 + e \vecb{A}(t)/\hbar} \, \hat{a}_{c, \vecb{k}_1 + e \vecb{A}(t)/\hbar} | \tilde \Psi(t) \rangle \right |^2,
\end{align}
where $\hat{a}_{v, \vecb{k}_1 + e \vecb{A}(t)/\hbar}$ is an annihilation operator corresponding to the valence state $u_{v, \vecb{k}_1 + e \vecb{A}(t)/\hbar}(\vecb{x})$, and $\hat{a}^{\dagger}_{c, \vecb{k}_1 + e \vecb{A}(t)/\hbar}$ is a creation operator corresponding to the conduction state $u_{c, \vecb{k}_1 + e \vecb{A}(t)/\hbar}(\vecb{x})$. Thus, the resulting state vector $\hat{a}^{\dagger}_{c, \vecb{k}_1 + e \vecb{A}(t)/\hbar} \hat{a}_{v, \vecb{k}_1 + e \vecb{A}(t)/\hbar}|\tilde \Phi_0 (t)\rangle$ corresponds to a Slater determinant formed by replacing $u_{v, \vecb{k}_1 + e \vecb{A}(t)/\hbar}(\vecb{x})$ with $u_{c, \vecb{k}_1 + e \vecb{A}(t)/\hbar}(\vecb{x})$ in Eq.~(\ref{appendix:eq:inst-gs-slater-solids}).

\begin{widetext}
The population of the one-particle one-hole state is then given by
\begin{align}
n_{vc, \vecb{k}_1 + e \vecb{A}(t)/\hbar} & =
\left | \int d\vecb{x} \, u^*_{c, \vecb{k}_1 + e \vecb{A}(t)/\hbar}(\vecb{x}) \, u_{\vecb{k}_1}(\vecb{x}, t) \right |^2
\times \left | \int d\vecb{x} \, u^*_{v, \vecb{k}_2 + e \vecb{A}(t)/\hbar}(\vecb{x}) \, u_{\vecb{k}_2}(\vecb{x}, t) \right |^2 \nonumber \\
& \times \cdots \times \left | \int d\vecb{x} \, u^*_{v, \vecb{k}_N + e \vecb{A}(t)/\hbar}(\vecb{x}) \, u_{\vecb{k}_N}(\vecb{x}, t) \right |^2.
\label{appendix:eq:1p1h-occupation-solids}
\end{align}

In a similar way, one may evaluate the population of a state using the number operator as
\begin{align}
n_{b, \vecb{k}_n + \frac{e\vecb{A}(t)}{\hbar}} = 
\langle \tilde{\Psi}(t)| \hat{a}^{\dagger}_{b, \vecb{k}_n + \frac{e\vecb{A}(t)}{\hbar}} 
\hat{a}_{b, \vecb{k}_n + \frac{e\vecb{A}(t)}{\hbar}} | \tilde{\Psi}(t)\rangle =
\left | \int d\vecb{x} \, u^*_{b, \vecb{k}_n + \frac{e\vecb{A}(t)}{\hbar}}(\vecb{x}) u_{\vecb{k}_n}(\vecb{x}, t) \right |^2.
\label{appendix:eq:band-occupation-single-band}
\end{align}

Hence, the instantaneous population of a band in the presence of a gauge field $\vecb{A}(t)$ can be evaluated by the overlap of the time-dependent Bloch function and the instantaneous eigenstate, or equivalently the wavevector-shifted state, also known as the Houston state. As discussed in Appendix~\ref{appendix:sec:ex-hydrogen}, this procedure is fundamentally equivalent to the evaluation of the time-dependent wavefunction and the field-free eigenstates in the length gauge.

One can straightforwardly extend Eq.~(\ref{appendix:eq:band-occupation-single-band}) for a multi-band system as
\begin{align}
n_{b, \vecb{k}_n + \frac{e\vecb{A}(t)}{\hbar}} = 
\langle \tilde{\Psi}(t)| \hat{a}^{\dagger}_{b, \vecb{k}_n + \frac{e\vecb{A}(t)}{\hbar}} 
\hat{a}_{b, \vecb{k}_n + \frac{e\vecb{A}(t)}{\hbar}} | \tilde{\Psi}(t)\rangle =
\sum_{b'} \left | \int d\vecb{x} \, u^*_{b, \vecb{k}_n + \frac{e\vecb{A}(t)}{\hbar}}(\vecb{x}) u_{b', \vecb{k}_n}(\vecb{x}, t) \right |^2.
\label{appendix:eq:band-occupation}
\end{align}

\end{widetext}

\section{Relaxation time approximation with the Houston states and Semiconductor Bloch equations \label{appendix:sec:houston-SBE}}

In this section, we revisit the expression for the instantaneous band population, Eq.~(\ref{appendix:eq:band-occupation}), computed with the Houston states from the perspective of the semiconductor Bloch equation. For this purpose, we consider the quantum master equation, Eq.~(\ref{eq:q-master}), with the relaxation time approximation, Eq.~(\ref{eq:relaxation-time-approx}). Since the band population, Eq.~(\ref{appendix:eq:band-occupation}), is computed with the Houston states, we set the Houton states as the reference states $\big|u^{\mathrm{Ref}}_{b\vecb{k}}(t)\big\rangle$ of the relaxation. The resulting relaxation operator is explicitly given by
\begin{widetext}
\begin{align}
\hat{D}\left [\rho_{\vecb{k}}(t) \right ] = 
& -\frac{1}{T_1}
\sum_{b} \big|u_{b,\vecb{k} + e\vecb{A}(t)/\hbar} \big\rangle 
\Big[\big\langle u_{b,\vecb{k} + e\vecb{A}(t)/\hbar} \big| \rho_{\vecb{k}}(t) \big|u_{b,\vecb{k} + e\vecb{A}(t)/\hbar} \big\rangle - \rho^{\mathrm{FD}}\left(\epsilon_{b,\vecb{k} + e\vecb{A}(t)/\hbar} \right)\Big]
\big\langle u_{b,\vecb{k} + e\vecb{A}(t)/\hbar} \big| \nonumber \\
& - \frac{1}{T_2}
\sum_{b \neq b'} \big|u_{b,\vecb{k} + e\vecb{A}(t)/\hbar} \big\rangle 
\big\langle u_{b,\vecb{k} + e\vecb{A}(t)/\hbar} \big| \rho_{\vecb{k}}(t) \big|u_{b',\vecb{k} + e\vecb{A}(t)/\hbar} \big\rangle 
\big\langle u_{b',\vecb{k} + e\vecb{A}(t)/\hbar}\big|.
\label{appendix:eq:relax-time-approx}
\end{align}

The quantum master equation, Eq.~(\ref{eq:q-master}), with the relaxation time approximation in the Houston basis, Eq.~(\ref{appendix:eq:relax-time-approx}), has been employed to investigate various phenomena~\cite{PhysRevB.99.214302,PhysRevB.99.224301,PhysRevB.106.024313}. However, its connection to the well-known semiconductor Bloch equation has not been thoroughly explored. Therefore, we revisit the derivation of the semiconductor Bloch equation from the quantum master equation, utilizing the relaxation operator constructed in the Houston basis.

For this purpose, we introduce the following quantity by allowing an arbitrary time dependence on $\vecb{k}$, defined as $\vecb{k}(t)$:
\begin{align}
\rho_{bb', \vecb{k}(t) + \frac{e\vecb{A}(t)}{\hbar}}(t) = \langle u_{b,\vecb{k}(t) + \frac{e\vecb{A}(t)}{\hbar}} |
\rho_{\vecb{k}(t)}(t) | u_{b',\vecb{k}(t) + \frac{e\vecb{A}(t)}{\hbar}} \rangle.
\end{align}
These are simply the matrix elements of the density matrix expressed in the Houston basis.

The equation of motion for $\rho_{bb', \vecb{k}(t) + \frac{e\vecb{A}(t)}{\hbar}}(t)$ is given as
\begin{align}
\frac{d}{dt} \rho_{bb', \vecb{k}(t) + \frac{e\vecb{A}(t)}{\hbar}}(t) &= 
\frac{d}{dt} \langle u_{b,\vecb{k}(t) + \frac{e\vecb{A}(t)}{\hbar}} |
\hat{\rho}_{\vecb{k}(t)}(t) | u_{b',\vecb{k}(t) + \frac{e\vecb{A}(t)}{\hbar}} \rangle \nonumber \\
&= \frac{\partial \rho_{bb', \vecb{k} + \frac{e\vecb{A}(t)}{\hbar}}(t)}{\partial \vecb{k}}\Big |_{\vecb{k}=\vecb{k}(t)}
\cdot \dot{\vecb{k}}(t)
+ \frac{\partial}{\partial \vecb{A}} \langle u_{b,\vecb{k}(t) + \frac{e\vecb{A}}{\hbar}} |
\hat{\rho}_{\vecb{k}(t)}(t) | u_{b',\vecb{k}(t) + \frac{e\vecb{A}}{\hbar}} \rangle \Big |_{\vecb{A}=\vecb{A}(t)} \cdot \dot{\vecb{A}}(t)  \nonumber \\
&\quad + \langle u_{b,\vecb{k}(t) + \frac{e\vecb{A}(t)}{\hbar}} |
\frac{d \hat{\rho}_{\vecb{k}}(t)}{dt} \Big |_{\vecb{k} = \vecb{k}(t)} | u_{b',\vecb{k}(t) + \frac{e\vecb{A}(t)}{\hbar}} \rangle \nonumber \\
&= \frac{\partial \rho_{bb', \vecb{k} + \frac{e\vecb{A}(t)}{\hbar}}}{\partial \vecb{k}} \Big |_{\vecb{k}=\vecb{k}(t)}
\cdot \dot{\vecb{k}}(t) \nonumber \\
&\quad + \left [
\left \langle \frac{\partial u_{b,\vecb{k} + \frac{e\vecb{A}(t)}{\hbar}}}{\partial \vecb{k}} \right |
\hat{\rho}_{\vecb{k}}(t) | u_{b',\vecb{k} + \frac{e\vecb{A}(t)}{\hbar}} \rangle 
+ \langle u_{b,\vecb{k} + \frac{e\vecb{A}(t)}{\hbar}} |
\hat{\rho}_{\vecb{k}}(t) \left | \frac{\partial u_{b',\vecb{k} + \frac{e\vecb{A}(t)}{\hbar}}}{\partial \vecb{k}} \right \rangle
\right ]_{\vecb{k}=\vecb{k}(t)} \cdot \frac{e \dot{\vecb{A}}(t)}{\hbar} \nonumber \\
&\quad + \langle u_{b,\vecb{k}(t) + \frac{e\vecb{A}(t)}{\hbar}} |
\left (
\frac{\left [ \hat{H}_{\vecb{k} + \frac{e\vecb{A}(t)}{\hbar}}, \hat{\rho}_{b\vecb{k}}(t) \right ]}{i\hbar} + \hat{D} \left [ \hat{\rho}_{b\vecb{k}}(t) \right ]
\right )_{\vecb{k} = \vecb{k}(t)} | u_{b',\vecb{k}(t) + \frac{e\vecb{A}(t)}{\hbar}} \rangle.
\label{appendix:eq:eom-density-matrix-elements-01}
\end{align}

For later convenience, we rewrite a quantity as
\begin{align}
& \left \langle \frac{\partial u_{b,\vecb k+e\vecb A(t)/\hbar}}{\partial \vecb k} \right |
\hat \rho_{\vecb k}(t)|u_{b',\vecb k+e\vecb A/\hbar}\rangle 
+ \langle u_{b,\vecb k+e\vecb A(t)/\hbar} |
\hat \rho_{\vecb k}(t) \left | \frac{\partial u_{b',\vecb k+e\vecb A(t)/\hbar}}{\partial \vecb k} \right \rangle \nonumber \\
=& \sum_a \Bigg[
\left \langle \frac{\partial u_{b,\vecb k+e\vecb A(t)/\hbar}}{\partial \vecb k} \right |u_{a,\vecb k + e\vecb A(t)/\hbar}\rangle \langle u_{a,\vecb k + e\vecb A(t)/\hbar} |
\hat \rho_{\vecb k}(t)|u_{b',\vecb k+e\vecb A/\hbar}\rangle \nonumber \\
&+ \langle u_{b,\vecb k+e\vecb A(t)/\hbar} |
\hat \rho_{\vecb k}(t) | u_{a,\vecb k + e\vecb A(t)/\hbar} \rangle \langle u_{a,\vecb k + e\vecb A(t)/\hbar} \left | \frac{\partial u_{b',\vecb k+e\vecb A(t)/\hbar}}{\partial \vecb k} \right \rangle
\Bigg] \nonumber \\
=& -i \sum_a \left[
\vecb{d}_{ab', \vecb k + e\vecb A(t)/\hbar} \rho_{ba, \vecb k + e\vecb A(t)/\hbar}
- \vecb{d}^*_{ab, \vecb k + e\vecb A(t)/\hbar} \rho_{ab', \vecb k + e\vecb A(t)/\hbar}
\right].
\end{align}
Here, the dipole matrix elements $\vecb{d}_{ab, \vecb k + e\vecb A(t)/\hbar}$ are defined as
\begin{align}
\vecb{d}_{ab, \vecb k + e\vecb A(t)/\hbar} = 
i \langle u_{a,\vecb k + e\vecb A(t)/\hbar} \left | \frac{\partial u_{b,\vecb k+e\vecb A(t)/\hbar}}{\partial \vecb k} \right \rangle.
\end{align}

Furthermore, we rewrite another quantity as
\begin{align}
&\langle u_{b,\vecb k+e\vecb A(t)/\hbar} |
\left(
\frac{\left[ \hat H_{\vecb k + e\vecb A(t)/\hbar}, \rho_{b\vecb k}(t)\right]}{i\hbar} + \hat D\left[\hat \rho_{b\vecb k}(t)\right]
\right) |u_{b',\vecb k+e\vecb A(t)/\hbar}\rangle \nonumber \\
&= \frac{1}{i\hbar} \left(\epsilon_{b,\vecb k+e\vecb A(t)/\hbar}
-\epsilon_{b',\vecb k+e\vecb A(t)/\hbar}
\right)\rho_{bb', \vecb k + e\vecb A(t)/\hbar} \nonumber \\
&-\frac{\delta_{bb'}}{T_1} \left(\rho_{bb,\vecb k + e\vecb A(t)/\hbar}(t) -
\rho^{\mathrm{eq}}_{b, \vecb k + e\vecb A(t)/\hbar}
\right)
-\frac{1-\delta_{bb'}}{T_2} \rho_{bb', \vecb k + e\vecb A(t)/\hbar}(t).
\end{align}

Hence, we can rewrite Eq.~(\ref{appendix:eq:eom-density-matrix-elements-01}) as
\begin{align}
& \frac{d}{dt} \rho_{bb', \vecb{k}(t) + \frac{e\vecb{A}(t)}{\hbar}}(t) = \frac{\partial \rho_{bb', \vecb{k} + \frac{e\vecb{A}(t)}{\hbar}}(t)}{\partial \vecb{k}}\Big|_{\vecb{k}=\vecb{k}(t)}
\cdot \dot{\vecb{k}}(t) \nonumber \\
& - i \frac{e\dot{\vecb{A}}(t)}{\hbar} \cdot \sum_{a}
 \left[
 \vecb{d}_{ab', \vecb{k}(t) + \frac{e\vecb{A}(t)}{\hbar}} \rho_{ba, \vecb{k}(t) + \frac{e\vecb{A}(t)}{\hbar}}(t) 
- \vecb{d}^*_{ab, \vecb{k}(t) + \frac{e\vecb{A}(t)}{\hbar}} \rho_{ab', \vecb{k}(t) + \frac{e\vecb{A}(t)}{\hbar}}(t)
\right] \nonumber \\
&+\frac{1}{i\hbar} \left(\epsilon_{b,\vecb k+e\vecb A(t)/\hbar}
-\epsilon_{b',\vecb k+e\vecb A(t)/\hbar}
\right)\rho_{bb', \vecb k + e\vecb A(t)/\hbar}(t) \nonumber \\
& - \frac{\delta_{bb'}}{T_1} \left(\rho_{bb, \vecb{k}(t) + \frac{e\vecb{A}(t)}{\hbar}}(t) - \rho^{eq}_{b, \vecb{k}(t) + \frac{e\vecb{A}(t)}{\hbar}}\right)
- \frac{1 - \delta_{bb'}}{T_2} \rho_{bb', \vecb{k}(t) + \frac{e\vecb{A}(t)}{\hbar}}(t).
\end{align}

We then set $\vecb{k}(t)$ to $\vecb{k} - \frac{e\vecb{A}(t)}{\hbar}$, and we obtain
\begin{align}
\frac{d}{dt} \rho_{bb', \vecb{k}}(t) & = - \frac{e\dot{\vecb{A}}(t)}{\hbar} \cdot \frac{\partial \rho_{bb', \vecb{k}}(t)}{\partial \vecb{k}} - i \frac{e\dot{\vecb{A}}(t)}{\hbar} \cdot \sum_{a}
 \left[
 \vecb{d}_{ab', \vecb{k}} \rho_{ba, \vecb{k}} (t)
- \vecb{d}^*_{ab, \vecb{k}} \rho_{ab', \vecb{k}}(t)
\right] \nonumber \\
&+\frac{1}{i\hbar} \left(\epsilon_{b\vecb k}
-\epsilon_{b'\vecb k}
\right)\rho_{bb', \vecb k}(t) \nonumber \\
& - \frac{\delta_{bb'}}{T_1} \left(\rho_{bb, \vecb{k}}(t) - \rho^{eq}_{b, \vecb{k}}\right)
- \frac{1 - \delta_{bb'}}{T_2} \rho_{bb', \vecb{k}}(t) \nonumber \\
& = \frac{e}{\hbar} \vecb{E}(t) \cdot \frac{\partial \rho_{bb', \vecb{k}}(t)}{\partial \vecb{k}}
+ i \frac{e}{\hbar} \vecb{E}(t) \cdot \sum_{a}
 \left[
 \vecb{d}_{ab', \vecb{k}} \rho_{ba, \vecb{k}} (t)
- \vecb{d}^*_{ab, \vecb{k}} \rho_{ab', \vecb{k}}(t)
\right] \nonumber \\
&+\frac{1}{i\hbar} \left(\epsilon_{b\vecb k}
-\epsilon_{b'\vecb k}
\right)\rho_{bb', \vecb k}(t) \nonumber \\
& - \frac{\delta_{bb'}}{T_1} \left(\rho_{bb, \vecb{k}}(t) - \rho^{eq}_{b, \vecb{k}}\right)
- \frac{1 - \delta_{bb'}}{T_2} \rho_{bb', \vecb{k}}(t).
\label{appendix:eq:sbe}
\end{align}
\end{widetext}

Here, Eq.~(\ref{appendix:eq:sbe}) refers to the semiconductor Bloch equation. We have demonstrated that the semiconductor Bloch equation is equivalent to the quantum master equation with the relaxation time approximation, based on the Houston basis. This result indicates that the band population, computed using the instantaneous eigenstates (equivalently, Houston states) and evaluated with Eq.~(\ref{appendix:eq:band-occupation-single-band}), is a physically reasonable choice, as it is consistent with the well-established relaxation time approximation in the semiconductor Bloch equation.

Each of the two equations of motion, Eq.~(\ref{appendix:eq:sbe}) and Eq.~(\ref{eq:q-master}) (with the Houston basis), has its own advantages and disadvantages for numerical calculations. The quantum master equation, Eq.~(\ref{eq:q-master}), using the Houston basis and the relaxation time approximation, can be solved stably in numerical simulations because it avoids quantities involving $\vecb{k}$-derivatives. However, it can be computationally demanding, as the relaxation operators require the calculation of instantaneous eigenstates at each time $t$. This necessitates the diagonalization of the Hamiltonian at every time step. 

In contrast, the semiconductor Bloch equation, Eq.~(\ref{appendix:eq:sbe}), can be solved more efficiently since it does not explicitly depend on momentum-shifted states and instead relies solely on quantities at $\vecb{k}$. However, it may be numerically unstable because it involves $\vecb{k}$-derivative terms. Therefore, when selecting the appropriate approach for practical applications, one must carefully weigh the characteristics of each equation of motion.

\section{Adiabatic evolution under a weak slowly-varying electric field \label{appendix:sec:adiabatic-evolution}}

Here, we prove that the polarized Houston states defined by Eq.~(\ref{eq:polarized-houston-states}) are the solutions of the time-dependent Schr\"odinger equation, Eq.~(\ref{eq:tdse}), in the limit of a weak and slowly-varying electric field, where the second order of the electric field $\vecb E(t)$ and the time derivative of the electric field $d\vecb E(t)/dt$ are negligible.

To this end, we consider a time-dependent wavefunction $|u_{b\vecb k}(t)\rangle$ that satisfies the following time-dependent Schr\"odinger equation:
\begin{align}
\left [i\hbar \frac{\partial}{\partial t} - \hat h_{\vecb k + e\vecb A(t)/\hbar} \right ]|u_{b\vecb k}(t)\rangle = 0.
\end{align}

We then expand $|u_{b\vecb k}(t)\rangle$ in terms of the polarized Houston states as
\begin{align}
|u_{b\vecb k}(t)\rangle = \sum_{b'} D_{b',b\vecb k}(t)|u^{PH}_{b'\vecb k}(t)\rangle,
\end{align}
where $D_{b',b\vecb k}(t)$ are the expansion coefficients.

\begin{widetext}
By multiplying a polarized Houston state from the left, we obtain the following relation:
\begin{align}
& \langle u^{PH}_{a\vecb k}(t) | \left [i\hbar \frac{\partial}{\partial t} - \hat h_{\vecb k + e\vecb A(t)/\hbar} \right ]|u_{b\vecb k}(t)\rangle =
\langle u^{PH}_{a\vecb k}(t) | \left [i\hbar \frac{\partial}{\partial t} - \hat h_{\vecb k + e\vecb A(t)/\hbar} \right ]\sum_{b'} D_{b',b\vecb k}(t)|u^{PH}_{b'\vecb k}(t)\rangle \nonumber \\
&=\left [i\hbar \dot{D}_{a,b\vecb k}(t) + \left (\epsilon^{P}_{a\vecb k}(t) -\hbar \dot{\gamma}^{P}_{a\vecb k}(t) \right )D_{a,b\vecb k}(t)  \right ] \nonumber \\
&+ \sum_{b'} D_{b',b\vecb k}(t)e^{i(\gamma^P_{b'\vecb k}(t)-\gamma^P_{a\vecb k}(t))}\exp \left [\frac{1}{i\hbar}\int^t_{-\infty}dt'~\epsilon^P_{b'\vecb k}(t')-\epsilon^P_{a\vecb k}(t') \right ]
\langle u^{P}_{a\vecb k}(t) | \left [i\hbar \frac{\partial}{\partial t} - \hat h_{\vecb k + e\vecb A(t)/\hbar} \right ]|u^{P}_{b'\vecb k}(t)\rangle = 0.
\label{eq:eom-for-pol-houston-expansion}
\end{align}

Here, we further evaluate the following quantity as
\begin{align}
& \langle u^{P}_{a\vecb k}(t) | \left [i\hbar \frac{\partial}{\partial t} - \hat h_{\vecb k + e\vecb A(t)/\hbar} \right ]|u^P_{b\vecb k}(t)\rangle 
=\sum_{a'b'}c^{P,*}_{aa'\vecb k}(t) \langle u^A_{a'\vecb k}(t)|\left [i\hbar \frac{\partial}{\partial t} - \hat h_{\vecb k + e\vecb A(t)/\hbar} \right ]c^{P}_{bb'\vecb k}(t)|u^A_{b'\vecb k}(t)\rangle \nonumber \\
&=i\hbar 
\sum_{a' b'} c^{P,*}_{aa'\vecb k}(t)i\hbar \dot{c}^{P,*}_{bb'\vecb k}(t)\delta_{a'b'}
+
\sum_{a'b'}c^{P,*}_{aa'\vecb k}(t)c^{P}_{bb'\vecb k}(t)  \langle u^A_{a'\vecb k}(t)|\left [i\hbar \frac{\partial}{\partial t} - \hat h_{\vecb k + e\vecb A(t)/\hbar} \right ]|u^A_{b'\vecb k}(t)\rangle \nonumber \\
&=i\hbar 
\sum_{a'} c^{P,*}_{aa'\vecb k}(t)i\hbar \dot{c}^{P,*}_{ba'\vecb k}(t)
-
\sum_{a'b'}c^{P,*}_{aa'\vecb k}(t)\left ( \mathcal{H}_{\mathrm{eff},\vecb{k}}(t) \right )_{a'b'} c^{P}_{bb'\vecb k}(t),
\label{eq:hamiltonian-matrix-elements-polarized-basis}
\end{align}
where $\left ( \mathcal{H}_{\mathrm{eff},\vecb{k}}(t) \right )_{a'b'}$ is the effective Hamiltonian defined in Eq.~(\ref{eq:effective-ham-matrix}). We note that the effective Hamiltonian $\mathcal{H}_{\mathrm{eff},\vecb{k}}(t)$ in Eq.~(\ref{eq:effective-tdse-adiabatic-basis}) can be described as
\begin{align}
\left ( \mathcal{H}_{\mathrm{eff},\vecb{k}}(t) \right )_{a'b'}
=\langle u^A_{a'\vecb k}(t)|\left [i\hbar \frac{\partial}{\partial t} - \hat h_{\vecb k + e\vecb A(t)/\hbar} \right ]|u^A_{b'\vecb k}(t)\rangle.
\end{align}

Since the eigenvectors $\vecb{c}^P_{b\vecb k}(t)$ are defined by Eq.~(\ref{eq:effective-ham-in-length-gauge}), Eq.~(\ref{eq:hamiltonian-matrix-elements-polarized-basis}) can be rewritten as
\begin{align}
& \langle u^{P}_{a\vecb k}(t) | \left [i\hbar \frac{\partial}{\partial t} - \hat h_{\vecb k + e\vecb A(t)/\hbar} \right ]|u^P_{b\vecb k}(t)\rangle  =
i\hbar 
\vecb{c}^{P,\dagger}_{a\vecb k}(t)
\dot{\vecb{c}}^P_{b\vecb k}(t)
-
\epsilon^P_{a\vecb k}(t)\delta_{ab}.
\label{eq:hamiltonian-matrix-elements-polarized-basis-mod}
\end{align}

By substituting Eq.~(\ref{eq:hamiltonian-matrix-elements-polarized-basis-mod}) into Eq.~(\ref{eq:eom-for-pol-houston-expansion}), the equation of motion for the expansion coefficients $D_{a,b\vecb k}(t)$ is given as
\begin{align}
i\hbar \dot{D}_{a,b\vecb k}(t) = -i\hbar 
\sum_{b' \neq a} D_{b',b\vecb k}(t)e^{i(\gamma^P_{b'\vecb k}(t)-\gamma^P_{a\vecb k}(t))}\exp \left [\frac{1}{i\hbar}\int^t_{-\infty}dt'~\epsilon^P_{b'\vecb k}(t')-\epsilon^P_{a\vecb k}(t')\right ]
\vecb{c}^{P,\dagger}_{a\vecb k}(t)
\dot{\vecb{c}}^P_{b'\vecb k}(t).
\end{align}

For further evaluation, we take the time derivative of both sides of Eq.~(\ref{eq:effective-ham-in-length-gauge}) and obtain the following:
\begin{align}
\dot{\mathcal{H}}_{\mathrm{eff},\vecb k}(t) \vecb{c}^P_{b\vecb k}(t)
+\mathcal{H}_{\mathrm{eff},\vecb k}(t) \dot{ \vecb{c}}^P_{b\vecb k}(t) = 
\dot{\epsilon}^P_{b\vecb k}(t) \vecb{c}^P_{b\vecb k}(t)
+\epsilon^P_{b\vecb k}(t) \dot{\vecb{c}}^P_{b\vecb k}(t).
\end{align}

By multiplying $\vecb{c}^{P,\dagger}_{a\vecb k}(t)$ from the left, the following expression can be obtained for $a\neq b$:
\begin{align}
\vecb{c}^{P,\dagger}_{a\vecb k}(t)\dot{ \vecb{c}}^P_{b\vecb k}(t) = 
\frac{\vecb{c}^{P,\dagger}_{a\vecb k}(t)\dot{\mathcal{H}}_{\mathrm{eff},\vecb k}(t) \vecb{c}^P_{b\vecb k}(t) }{\epsilon^P_{a\vecb k}(t)-\epsilon^P_{b\vecb k}(t)}.
\end{align}

By taking the time derivative of Eq.~(\ref{eq:effective-ham-matrix}), the following expression is derived:
\begin{align}
\frac{d}{dt} \left ( \mathcal{H}_{\mathrm{eff},\vecb{k}}(t) \right )_{bb'} &= -\frac{e}{\hbar}\frac{\partial \epsilon_{b, \vecb{k} + e\vecb{A}(t)/\hbar}}{\partial \vecb k}\cdot \vecb E(t) \delta_{bb'} +\frac{d}{dt}\left [ i \left (1 - \delta_{bb'} \right )  \frac{e\vecb{E}(t)}{\hbar} \cdot \langle u^{\mathrm{A}}_{b\vecb{k}}(\vecb{r},t) | \frac{\partial}{\partial \vecb{k}} | u^{\mathrm{A}}_{b'\vecb{k}}(\vecb{r},t) \rangle \right ].
\label{eq:effective-ham-matrix-time-derivative}
\end{align}
Here, the second term on the right-hand side consists of terms of the first order of $\dot{\vecb E}(t)$ or the second order of $\vecb E(t)$. Hence, by ignoring such terms in the weak and slowly-varying electric field limit, the matrix elements can be approximated as
\begin{align}
\frac{d}{dt} \left ( \mathcal{H}_{\mathrm{eff},\vecb{k}}(t) \right )_{bb'} 
= -\frac{e}{\hbar}\frac{\partial \epsilon_{b, \vecb{k} + e\vecb{A}(t)/\hbar}}{\partial \vecb k}\cdot \vecb E(t) \delta_{bb'} +\mathcal{O}\left ( \mathrm{max}\left (|\dot{\vecb {E}}(t)|,  |\vecb {E}(t)|^2 \right ) \right ).  
\label{eq:effective-ham-matrix-time-derivative-approx}
\end{align}

For further evaluation, we decompose $\vecb{c}^P_{b\vecb k}(t)$ into the 0th order of $\vecb E(t)$ and others as
\begin{align}
\vecb{c}^P_{b\vecb k}(t) = \vecb{c}^{P,\mathrm{0th}}_{b\vecb k}(t) + \delta \vecb{c}^P_{b\vecb k}(t).
\end{align}
Furthermore, the matrix elements of Eq.~(\ref{eq:effective-ham-matrix}) are diagonal up to the 0th order of $\vecb E(t)$. Hence, one may naturally express the elements of $\vecb{c}^{P,\mathrm{0th}}_{b\vecb k}(t)$ as
\begin{align}
c^{P,\mathrm{0th}}_{bb'\vecb k}(t) = \delta_{bb'}.
\end{align}

With these expressions, the following quantity for ($a\neq b$) is further evaluated as
\begin{align}
\vecb{c}^{P,\dagger}_{a\vecb k}(t)\dot{\mathcal{H}}_{\mathrm{eff},\vecb k}(t) \vecb{c}^P_{b\vecb k}(t)
&= -\frac{e}{\hbar} \sum_{a'b'}c^{P,*}_{aa'\vecb k}(t)
\frac{\partial \epsilon_{b, \vecb{k} + e\vecb{A}(t)/\hbar}}{\partial \vecb k}\cdot \vecb E(t) \delta_{a'b'}
c^P_{bb'\vecb k}(t) + \mathcal{O}\left ( \mathrm{max}\left (|\dot{\vecb {E}}(t)|,  |\vecb {E}(t)|^2 \right ) \right ) \nonumber \\
&= -\frac{e}{\hbar} \sum_{a'b'}c^{P,\mathrm{0th},*}_{aa'\vecb k}(t)
\frac{\partial \epsilon_{b, \vecb{k} + e\vecb{A}(t)/\hbar}}{\partial \vecb k}\cdot \vecb E(t) \delta_{a'b'}
c^{P,\mathrm{0th}}_{bb'\vecb k}(t)
+\mathcal{O}\left ( \mathrm{max}\left (|\dot{\vecb {E}}(t)|,  |\vecb {E}(t)|^2 \right ) \right ) \nonumber \\
&=\mathcal{O}\left ( \mathrm{max}\left (|\dot{\vecb {E}}(t)|,  |\vecb {E}(t)|^2 \right ) \right ).
\end{align}

\end{widetext}

Likewise, the following quantity for ($a\neq b$) is further evaluated as
\begin{align}
\vecb{c}^{P,\dagger}_{a\vecb k}(t)\dot{ \vecb{c}}^P_{b\vecb k}(t) = 
\mathcal{O}\left ( \mathrm{max}\left (|\dot{\vecb {E}}(t)|,  |\vecb {E}(t)|^2 \right ) \right ),
\end{align}
and
\begin{align}
\dot{D}_{a,b\vecb k}(t) = \mathcal{O}\left ( \mathrm{max}\left (|\dot{\vecb {E}}(t)|,  |\vecb {E}(t)|^2 \right ) \right ).
\label{eq:pol-houston-expansion-coeff-linear-slow-limit}
\end{align}

The final expression of Eq.~(\ref{eq:pol-houston-expansion-coeff-linear-slow-limit}) indicates that the expansion coefficients $D_{a,b\vecb k}(t)$ become constant when the first order of $\dot{\vecb E}(t)$ and the second order of $\vecb{E}(t)$ are ignored. In other words, the polarized Houston states are accurate solutions of the time-dependent Schr\"odinger equation [Eq.~(\ref{eq:tdse})] in the limit of a weak and slowly varying electric field.

The above analysis implies that the polarized Houston states are exact solutions of the time-dependent Schr\"odinger equation in the static and linear response limits. Hence, when the relaxation operator in Eq.~(\ref{eq:relaxation-time-approx}) is constructed with the polarized Houston states, the induced dynamics are not affected by the relaxation operator. As a result, the spurious current induced in the relaxation time approximation with the Houston basis is fully removed by utilizing the polarized Houston basis.

\bibliography{ref}

\end{document}